\documentclass[12pt]{article}
\usepackage{amsmath,amssymb,amsthm,bm,mathabx,graphicx}

\usepackage{times}  
\usepackage{multirow}  
\usepackage{color}
\usepackage{xr} 
\usepackage{algorithm}
\usepackage{algpseudocode}

\usepackage{graphicx,subfig,psfrag,epsf}
\usepackage{multirow}
\usepackage{enumerate,array,booktabs}
\usepackage{natbib}
\usepackage[hidelinks]{hyperref}
\usepackage{enumitem}
\usepackage{caption}
\usepackage{rotating}  
\usepackage{amsfonts}
\usepackage{dsfont} 
\usepackage{pdflscape}
\usepackage{alltt}
\usepackage[T1]{fontenc} 
\usepackage{pifont} 
\usepackage{booktabs}
\IfFileExists{upquote.sty}{\usepackage{upquote}}{}

\usepackage{setspace}
\makeatletter
\def\singlespace{\def\baselinestretch{1}\@normalsize}

\newenvironment{newproof}[1]{ \paragraph{Proof of {#1}:}}{\hfill$\Box$\vspace{1em}}
\newenvironment{keywords}{ \paragraph{Keywords:}}

\newcommand{\vs}{\bm{s}}

\newcommand{\vx}{\bm{x}}

\newcommand{\vH}{\bm{H}}
\newcommand{\vI}{\bm{I}}

\newcommand{\vQ}{\bm{Q}}

\newcommand{\vR}{\bm{R}}
\newcommand{\vT}{\bm{T}}

\newcommand{\vX}{\bm{X}}

\newcommand{\vD}{\bm{D}}

\newcommand{\indicator}{\mbox{I}}
\newcommand{\vnull}{\bm{0}}

\newcommand{\bqa}{\begin{eqnarray*}}
	\newcommand{\eqa}{\end{eqnarray*}}
\newcommand{\bqan}{\begin{eqnarray}}
\newcommand{\eqan}{\end{eqnarray}}

\newcommand{\vdelta}{\bm{\delta}}

\newcommand{\vbeta}{\bm{\beta}}

\newcommand{\vgamma}{\bm{\gamma}}

\newcommand{\veta}{\bm{\eta}}

\newcommand{\vtheta}{\bm{\theta}}
\newcommand{\vxi}{\bm{\xi}}

\newcommand{\vpsi}{\bm{\psi}}

\newcommand{\bay}{\begin{array}}
	\newcommand{\eay}{\end{array}}

\newcommand{\vxw}{\widetilde{\vx}}
\newcommand{\vdeltaw}{\widetilde{\vdelta}}
\newcommand{\vbetaw}{\widetilde{\vbeta}}
\newcommand{\vbetah}{\widehat{\vbeta}}
\newcommand{\vsw}{\widetilde{\vs}}

\newcommand{\bbP}{\mathbb{P}}
\newcommand{\bbR}{\mathbb{R}}
\newcommand{\bbB}{\mathbb{B}}
\newcommand{\bbD}{\mathbb{D}}

\newcommand{\E}{\mbox{E}}
\newcommand{\Var}{\mbox{Var}}

\newtheorem{theorem}{Theorem}
\newtheorem{lemma}{Lemma}

\newtheorem{remark}{Remark}

\newcommand{\ra}{\rightarrow}
\newcommand{\rap}{ \xrightarrow{p}}
\newcommand{\rad}{ \xrightarrow{d}}

\newcommand{\oprw}{o_{p_{rw}}(1)}
\newcommand{\opr}{o_{p_{r}}(1)}
\newcommand{\opw}{o_{p_{w}}(1)}
\newcommand{\pxp}{}

\newcommand{\CR}{{Coverage Rate}}
\newcommand{\AL}{{Average Length}}

\def\boxit#1{\vbox{\hrule\hbox{\vrule\kern6pt
			\vbox{\kern6pt#1\kern6pt}\kern6pt\vrule}\hrule}}

\title{Resampling-based Confidence Intervals for Model-free Robust Inference on Optimal Treatment Regimes}

\author{Yunan Wu$^{*}$ and 
	Lan Wang$^{**}$\\
	\small $^{*}$ School of Statistics, University of Minnesota, Minneapolis, MN 55455, U.S.A.\\
\small$^{**}$ Department of Management Science, University of Miami, Coral Gables, FL 33146\\
\small $^{*}$ \textit{email:} wuxx1375@umn.edu\\
\small $^{**}$ \textit{email:} lanwang@mbs.miami.edu}
\date{}

\begin{document}
 
 	\maketitle

\begin{abstract}
We propose a new procedure for inference on optimal treatment regimes in the 
model-free setting, which does not require to specify an outcome regression model. 
Existing model-free estimators for optimal treatment regimes are usually not suitable for the
purpose of inference, because they either have nonstandard asymptotic distributions
or do not necessarily guarantee consistent estimation of the parameter indexing the Bayes rule
due to the use of surrogate loss. 
We first study a smoothed robust estimator that
directly targets the parameter corresponding to the Bayes decision rule for optimal treatment regimes estimation.
This estimator is shown to have an asymptotic normal distribution. 
Furthermore, we verify that a resampling procedure provides asymptotically accurate inference for both the parameter indexing the optimal treatment regime and the optimal value function. A new algorithm is developed to calculate the proposed estimator with substantially improved speed and stability. Numerical results demonstrate the satisfactory performance of the new methods.
\end{abstract}


\section{Introduction}\label{sec:intro}
Applications in medicine, public policy, internet marketing and other scientific areas often require estimating an individualized treatment rule (or regime, policy) to maximize the potential benefit. For example, \citet{Gail} and  \citet{zhang2012} observed that younger patients with primary operable breast cancer and lower PR levels are likely to benefit more from the treatment L-phenylalanine mustard and 5-fluorouracil (PF) rather than from PF plus tamoxifen (PFT). 
Several successful estimation strategies have been developed, including Q-learning \citep{watkins1992, Murphy05, chak2010, qian2011, Song2015}, A-learning \citep{robins2000, Murphy2003, murphy2005experimental, moodie2010, Shi2018}, model-free methods \citep{robins2008, Orellan10, zhang2012, zhao2012estimating, zhao2015, athey2017, Linn17, zhou2017residual, zhu2017, Wang2018, qi2018d, Lou2018optimal}, tree or list-based methods \citep{Laber2015tree, Cui2017tree, zhu2017, zhang2018estimation}, targeted learning ensembles approach \citep{Diaz2018}, among others. 


This paper focuses on inference for optimal treatment regimes. In practice, it is often desirable to have an interpretable treatment regime. Here, we focus on the popular class of index rules, given by $\bbD=\{\indicator(\vx^T\vbeta> 0): \vbeta\in \bbB\}$,  
where $\indicator(\cdot)$ is the indicator function and $\mathbb{B}$ is a compact subset of $ \mathbb{R}^p$. 
We consider two important inference targets: one is the parameter  $\vbeta_0$ indexing the theoretically optimal treatment regime and the other is the theoretically optimal value function $V(\vbeta_0)$. The former inference problem helps understand
the importance of different predictors on making an optimal decision, while the latter 
aims to quantify the maximally achievable expected performance which can be used as a gold standard to 
evaluate alternative treatment regimes.

Although there exists a rich literature on estimation,  the associated inference problem has not been studied until recently.  For Q-learning, several inference methods have been investigated.
\cite{laber2010} proposed a novel locally consistent adaptive confidence interval for  $\vbeta_0$,
\cite{chakraborty2013} proposed a practically convenient adaptive $m$-out-of-$n$ bootstrap for inference on  $\vbeta_0$, 
\cite{Chakraborty2014} introduced a double bootstrap approach for inference for $V(\vbeta_0)$, 
 \citet{Song2015} considered inference for $\vbeta_0$ based on the asymptotic distribution theory for 
penalized Q-learning.  Recently, \citet{Jeng2018} developed Lasso-based procedure for inference on $\vbeta_0$ in the A-learning framework. 
However, accurate inference based on Q-learning and A-learning needs reliable model specification. \citet{Luedtke2016} developed interesting theory for inference for $V(\vbeta_0)$ under exceptional laws. Their approach requires to estimate the conditional treatment effect either based on a working model or in a completely nonparametric fashion.

Different from current state-of-the-art methods which are mostly model-based,
we aim to develop a model-free approach for making inference for both 
$\vbeta_0$ and $V(\vbeta_0)$.
This would be useful to alleviate the sensitivity of inference with respect to the underlying generative model, the specification of which is often challenging in real data analysis.
It is known that the parameter indexing the optimal treatment regime $\vbeta_0$ corresponds to the parameter of the Bayes
rule of a weighted classification problem \citep{qian2011,zhang2012,zhao2012estimating}.  
A substantial challenge in inference for $\vbeta_0$  lies in the nonsmoothness of the
decision function. A popular approach is to replaces the 0-1 loss by a computationally
convenient surrogate loss such as the hinge loss \citep{zhao2012estimating, zhou2017residual,Lou2018optimal} or the logistic loss 
\citep{jiang2019entropy}. However, existing theory
(e.g., Fisher consistency, generalization error bound) that 
justifies the use of the surrogate loss is usually derived when the
form of the decision rule is unconstrained and approximated in a 
reproducible kernel Hilbert space.
There is no guarantee that when we consider the class of decision rules $\bbD$,
use of surrogate loss still leads to a decision function whose sign matches $\mbox{sign}(\vx^T\vbeta_0)$, 
see \citet{Lin2002}.  On the other hand, robust estimator \citep{zhang2012} that directly estimates the Bayes rule
has a cubic root convergence rate and a nonnormal limiting distribution, as recently revealed in 
\cite{Wang2018}. Inference is challenging due to the nonstandard asymptotics as naive bootstrap procedure is not consistent.
\citet{Goldberg2014} proposed a SoftMax Q-learning approach to alleviate the nonsmoothness
problem in Q-learning but have not explore the associated inference theory.

This paper first proposes a smoothed model-free estimator for the optimal treatment regime and introduce a proximal algorithm which substantially improves both the computational speed and the accuracy. We prove that the smoothed robust estimator has an asymptotic normal distribution and converges to $\vbeta_0$ with a rate that can be made arbitrarily close to $n^{-1/2}$. We then rigorously justify the validity of a resampling approach for inference.  
Our study focuses on randomized trials. Extension to observational study is discussed in Section~\ref{sec:diss}.

The remaining of the paper is organized as follows. Section~\ref{sec:method} introduces the new method and algorithm.
Section~\ref{sec:theory} carefully studies the statistical properties for estimation and inference. Section~\ref{sec:simulate} reports the results from Monte Carlo simulations.
Section~\ref{sec:realdata} analyzes a clinical data set from the Childhood Adenotonsillectomy Trial (CHAT). Section~\ref{sec:diss} concludes with some discussions. 
The appendix gives the technical assumptions and presents several useful lemmas.
The  supplemental file contains additional numerical results and detailed technical derivations.

\section{Proposed Methods}\label{sec:method}
\subsection{Problem Setup}\label{setup}
Let $A$ be a binary variable (0 or 1) denoting the treatment.  For each subject, we observe a vector of covariates
$\vx\in\bbR^p$ and an outcome $Y\in \bbR$.   Without loss of generality, we assume that larger outcome is preferred.
To evaluate the treatment effect, we adopt the potential or counterfactual outcome framework \citep{Neyman1990, Rubin1978} for causal inference. 
Let $Y^*_1$ and $Y^*_0$ be the potential outcome had the subject received treatment 1 and 0, respectively. 
In reality, we observe either $Y^*_1$ or $Y^*_0$, but never both. It is assumed that the observed outcome is the potential outcome corresponding to the treatment the subject actually receives 
(consistency assumption in causal inference), that is
$Y = Y^*_1 A + Y^*_0 (1-A)$.
Assume $A$ and $\{Y^*_0, Y^*_1\}$ are independent conditional on $\vx$, that is, no unmeasured confounding. 
In addition, we assume that the stable unit treatment value assumption \citep{Rubin86} and the positivity assumption are both satisfied, where the former requires a subject's outcome from receiving a treatment is not influenced by the treatment received by other subjects
and the latter requires that $0<P(A=a|\vx)<1$, $\forall \ \vx$, almost surely. 

An individualized treatment rule or a treatment regime, denoted by $d(\vx)$, is a mapping 
from the space of covariates to the set of treatment options $\{0, 1\}$. 
Let $Y^*(d)$ be the potential outcome had a subject with covariates $\vx$ received the treatment assigned by $d(\vx)$.  
We have
\bqan \label{cat1}
 Y^*(d)=Y^*_1 d(\vx)+Y^*_0\{1-d(\vx)\}.
 \eqan
Given a collection $\mathbb{D}$ of treatment regimes, 
the optimal treatment regime $\arg \max_{d\in \mathbb{D}}\E(Y^*(d))$ leads to the maximal average outcome if being implemented in the population.

For a given $\vbeta\in \bbB$, we sometimes write the corresponding treatment regime 
$\indicator(\vx^T\vbeta> 0)$ as $d_{\vbeta}(\vx)$ or $d_{\vbeta}$ for simplicity. 
The value function $V(\vbeta)=\E \{Y^*(d_{\vbeta})\}$ measures the effectiveness of 
the treatment regime $d_{\vbeta}$.
We are interested in estimating the parameter indexing the optimal rule 
\bqan\label{truebeta}
\vbeta_0=\arg\max_{ \vbeta\in \bbB}V(\vbeta).
\eqan

For identifiability, we assume that there exists a covariate with a nonzero coefficient whose conditional distribution given the other covariates is absolutely continuous and its coefficient is normalized to have absolute value one. 
Without loss of generality (one can rearrange the labels of the predictors),
we assume $x_1$ is a predictor that satisfies the condition.
We write $\vbeta = (\beta_1,\vbetaw^T)^T\in\bbR^p$. Correspondingly, we write $\vx = (x_1, \vxw^T)^T$. 
More discussions on alternative identifiability condition can be found in Section 6.2.


\subsection{Challenges of inference based on existing robust estimators}\label{challege}
\citet{qian2011}, \citet{zhang2012}, \citet{zhao2012estimating}, among other, observed that optimal treatment regime estimation can
be reformulated as a weighted classification problem. Specifically,
the value function $V(\vbeta)$ can be equivalently expressed as
\bqan\label{classify}
V(\vbeta)=\E\Big[\frac{Y}{\pi(A, \vx)}I\big\{A= d_{\vbeta}(\vx)\big\}\Big],
\eqan
where $\pi(A, \vx)=P(A=1|\vx)$ is the propensity score of the treatment and is equal to $0.5$ in a randomized trial. Expression (\ref{classify}) is the foundation for robust or policy-search estimators for optimal treatment regime, which aim to alleviate the practical difficulty of specifying a reliable generative regression model.

A robust estimator can be obtained by directly maximizing an unbiased sample estimator of the expectation in (\ref{classify}), which was the approach in  \citet{zhang2012}. In a randomized trial, based on the observed data $\{(\vx_i, Y_i, A_i), i=1,\ldots,n\}$, which are independent copies of $(\vx, Y, A)$, $V(\vbeta)$ can be consistently estimated by its sample analog 
\bqan\label{Vn}
V_n(\vbeta)= \frac{2}{n}\sum_{i=1}^n\{A_i\indicator(\vx_i^T\vbeta> 0)+(1-A_i)\indicator(\vx_i^T\vbeta\leq 0)\}Y_i.
\eqan 
Leaving out the terms in $V_n(\vbeta)$ that do not depend on $\vbeta$, we can estimate $\vbeta_0$ by
\bqan\label{lion}
\arg\max_{ \vbeta\in \bbB}M_n(\vbeta)=\arg\max_{ \vbeta\in \bbB}  \frac{2}{n}\sum_{i=1}^n(2A_i-1)\indicator(\vx_i^T\vbeta> 0)Y_i.
\eqan
However, as revealed in \cite{Wang2018} such a direct estimator for the Bayes rule belongs to a class of nonstandard $M$ estimators. It converges at a cubic-root rate to a nonnormal limiting distribution that is characterized by the maximizer of a centered Gaussian process with a parabolic drift. The nonstandard asymptotics is a consequence of the so-called {\it sharp-edge effect} \citep{KimPollard90}.
Inference based on this approach is challenging due to the nonstandard asymptotics as the naive bootstrap procedure is not consistent.
The smoothed estimator we propose alleviates the sharp-edge effect caused by the indicator function and leads to faster convergence rate.

\subsection{Smoothed Model-free Inference for Optimal Treatment Regime}\label{dragon}
To facilitate inference, we study an alternative estimator which can be considered as a compromise between the two robust estimation approaches described in Section~\ref{challege}. For clarity of presentation, we assume that the data are collected from a randomized trial.
Instead of replacing the indicator function with the hinge loss function, we replace it with a smoothed approximation. Formally, we estimate $\vbeta_0$ by
\begin{align}
\label{smoothed}
\vbetah_n=\arg\max_{ \vbeta\in \bbB} \widetilde{M}_n(\vbeta) =
 \arg\max_{\vbeta\in \bbB}\frac{2}{n}\sum_{i=1}^n(2A_i-1)K\Big(\frac{\vx_i^T\vbeta}{h_n}\Big)Y_i,
\end{align}
where $K(\cdot)$ is a smoothed approximation to the indicator function, and $h_n$ is a sequence of smoothing parameter that goes to zero as $n\ra \infty$.
The function  $K(\cdot)$ is required to satisfy some general regularity conditions given in the Appendix, see also Remark 1
in Section 3.1.

The motivation for the above new estimator is three-fold. 
First, as $h_n$ goes to zero at an appropriate rate, the parameter indexing the optimal
treatment regime or the Bayes rule can be estimated at a rate arbitrarily close to $n^{-1/2}$, see Section \ref{Theory1}.
Second, smoothing the indicator function circumvents
the aforementioned nonstandard asymptotics and would lead to a feasible bootstrap inference procedure with theoretical
guarantee, see Section \ref{Theory2}. Third, it also alleviates the computational challenge due to nonsmoothness, see Section \ref{algorithm} for a new efficient algorithm. 



For inference, we apply a resampling technique called ``weighted bootstrap" which assigns independent and identically distributed positive random weights to each observation. This resampling scheme was proposed in  \citet{Rubin1981}. \citet{WB} provided a comprehensive introduction, see also \citet{Ma2005} and \citet{ChengHuang} for recent interesting developments. The bootstrapped estimate of the smoothed robust estimator is defined as
\begin{align}
\label{boots}
\vbetah^*_n=\arg\max_{ \vbeta\in \bbB} \widetilde{M}^*_n(\vbeta) =\arg\max_{ \vbeta\in \bbB}\frac{2}{n}\sum_{i=1}^n r_{i} 
(2A_i-1) K\Big(\frac{\vx_i^T\vbeta}{h_n}\Big)Y_i,
\end{align}
where $r_{1},...,r_{n}$ are random weights satisfying conditions given in Section 3.2. 
To evaluate the distribution of $\vbetah^*_n$ in practice, we repeatedly generate independent samples of random weights. 
Following notation introduced earlier, let $\vbetah^*_n=(\widehat{\beta}^*_{n1}, \widetilde{\vbeta}^{*T}_n)^T$, where $|\widehat{\beta}^*_{n1}|=1$ and $\widetilde{\vbeta}^{*}_n=(\widehat{\beta}^*_{n2}, \ldots, \widehat{\beta}^*_{np})^T$. For $j=2,\ldots,p$, let $\xi_j^{*(\alpha/2)}$ and $\xi_j^{*(1-\alpha/2)}$ be the $(\alpha/2)$-th and $(1-\alpha/2)$-th quantile of the bootstrap distribution of   $(nh_n)^{1/2}(\widetilde{\vbeta}_j^*-\widetilde{\vbeta}_j)$, respectively, where $\alpha$ is a small positive number. We can estimate $\xi_j^{*(\alpha/2)}$ and $\xi_j^{*(1-\alpha/2)}$ from a large number of bootstrap samples. An asymptotic $100(1-\alpha)\%$ bootstrap confidence interval for $\beta_{0j}$,   $j=2,\ldots, p$, is given by
\bqan\label{b1}
\big\{\widetilde{\vbeta}_j-(nh_n)^{-1/2}\xi_j^{*(1-\alpha/2)}, \widetilde{\vbeta}_j-(nh_n)^{-1/2}\xi_j^{*(\alpha/2)}\big\}.
\eqan

Next, we consider inference for the optimal value.  Define
\bqan\label{Vnstar}
V_n^*(\vbeta)= 
\frac{2}{n}\sum_{i=1}^nr_i\{A_i\indicator(\vx_i^T\vbeta> 0)+(1-A_i)\indicator(\vx_i^T\vbeta\leq 0)\}Y_i.
\eqan 
Note that $V_n^*(\vbeta)$ can be considered as a perturbed version of the $V_n$ defined in (\ref{Vn}). Let  $d^{*(\alpha/2)}$ and $d^{*(1-\alpha/2)}$ be the $(\alpha/2)$-th and $(1-\alpha/2)$-th quantile of the bootstrap distribution of   $n^{1/2}\{V_n^*(\vbetah_n)-V_n(\vbetah_n)\}$, respectively. An asymptotic $100(1-\alpha)\%$ bootstrap confidence interval for $V(\vbeta_0)$ is 
\bqan\label{b2}
\big\{V_n(\vbetah_n)-n^{-1/2}d^{*(1-\alpha/2)}, V_n(\vbetah_n)-n^{-1/2}d^{*(\alpha/2)}\big\}.
\eqan


\subsection{A Proximal Algorithm}\label{algorithm}
The smoothed robust estimator largely alleviates the computational challenge due to the nonsmooth indicator function. However, the objective function is still a nonconvex function of the parameter. Such nonconvexity is inherent to robust estimation of optimal treatment regime \citep{qian2011}. We employ a proximal gradient descent algorithm, originally proposed in \citet{Nesterov}, which applies to a large class of nonconvex problems. In our setting, this algorithm substantially improves the computational speed and can accommodate high-dimensional covariates.

 Consider an optimization problem with an objective function $\Phi(\vbeta)$. \citet{Nesterov} assumes that $\Phi(\vbeta)$ has the  decomposition $ \Phi(\vbeta) = f(\vbeta) + \Psi (\vbeta),$
over a convex set $Q$, where $f$ is a differentiable function but not necessarily convex, and $\Psi$ is closed and convex on $Q$. 
In our setting, we take $- \widetilde{M}_n(\vbeta)$ as the $f$ function, and set $ \Psi (\vbeta)\equiv 0$.
Following \citet{Nesterov}, we generate a sequence of iterates $\{\vbeta^{(t)}, t = 0,1,2,...\}$ such that
\begin{align*}
\vbeta^{(t)} = \arg\min\limits_{\vbeta  \in \bbB} \Big\{&-\widetilde{M}_n(\vbeta^{(t-1)}) - \big\langle \nabla \widetilde{M}_n(\vbeta^{(t-1)}), \vbeta-\vbeta^{(t-1)} \big\rangle + \alpha_t\big|\big|\vbeta-\vbeta^{(t-1)}\big|\big|^2 + \Psi(\vbeta)\Big\}\\
=\arg\min\limits_{\vbeta  \in \bbB} \Big\{& -\frac{2}{n}\sum_{i=1}^n(2A_i-1)K'\Big(\frac{\vx_i^T\vbeta^{(t-1)}}{h_n}\Big)\frac{\vx_i^T(\vbeta-\vbeta^{t-1})}{h_n}Y_i  + \alpha_t\big|\big|\vbeta-\vbeta^{(t-1)}\big|\big|^2\Big\},
\end{align*}
where $\langle \cdot, \cdot \rangle$ denotes the inner product between two vectors.
Observe that the above minimization problem has a closed-form solution 
\bqa
\vbeta^{(t)}  = \vbeta^{(t-1)} + (n\alpha_t)^{-1}\sum_{i=1}^n(2A_i-1)K'\Big(\frac{\vx_i^T\vbeta^{(t-1)}}{h_n}\Big)\frac{\vx_i}{h_n}Y_i.
\eqa
Hence the algorithm can be updated efficiently. The algorithm stops  when the following criterion is met: 
\begin{align*} 
\widetilde{M}_n(\vbeta^{(t)})< \widetilde{M}_n(\vbeta^{(t-1)}) + \big\langle \nabla \widetilde{M}_n(\vbeta^{(t-1)}), \vbeta^{(t)}-\vbeta^{(t-1)} \big\rangle - \alpha_t\big|\big|\vbeta^{(t)}-\vbeta^{(t-1)}\big|\big|^2, 
\end{align*}
where $\alpha_t$ is a sequence of small positive numbers.
To choose $\alpha_t$, inspired by \citet{Fan2018}, we employ an expanding series, which ensures that the stepsize diminishes during the update process. Details for this algorithm is provided in the supplementary material.

It is worth emphasizing that this algorithm can be easily adapted to the high-dimensional setting 
by taking $\Psi(\vbeta)$ as a regularization function, such as  the $L_1$ penalty function. 

\section{Statistical Properties}\label{sec:theory}
\subsection{Consistency and Asymptotic Normality of the Smoothed Estimator} \label{Theory1}
To lay the foundation for inference, we first present the statistical properties 
of the smoothed robust estimator $\vbetah_n$ defined in (\ref{smoothed}).
All the regularity conditions are summarized in the Appendix.
Theorem~\ref{Th1} below shows that $\vbetah_n$ is consistent for the parameter indexing the optimal treatment regime. 
Comparing with the asymptotic normality result in Theorem~\ref{Th2}, the consistency requires very mild conditions and serves as a precursor step for proving asymptotic normality.  See Section S2 of the online supplementary material for the proofs of Theorem~\ref{Th1} and Theorem~\ref{Th2}.

\begin{theorem}
	\label{Th1}
	Under \ref{A1} - \ref{A3} and assume $K(\cdot)$ satisfies \ref{K1}, then  $\vbetah_n = \vbeta_0+o_p(1)$. 
\end{theorem}
 
Recall that for identification, we write $\vbeta_0 = (\beta_{01},\vbetaw_0^T)^T\in\bbR^p$ where $|\beta_{01}| = 1$. 
Similarly, we write $\vbetah_n = (\widehat{\beta}_{n1},\vbetaw_n^T)^T\in\bbR^p$ where $|\widehat{\beta}_{n1}| = 1$.
With the above consistency result, we have $P(\widehat{\beta}_{n1}=\beta_{01})\ra 1$ as $n\ra \infty$.
In the following, we focus on studying the asymptotic distribution of $\vbetaw_n$.
To this end, we introduce some additional notations. 
Define $ S(z,\vxw) = \E(Y^*_1-Y^*_0|z,\vxw)$,
where $z=\vx^T\vbeta_0$. Note that there is a one-to-one transformation between $(z, \vxw)$ and  $\vx = (x_1, \vxw^T)^T$. 
Hence, $ S(z,\vxw)$ is a measure of the conditional treatment effect.
Let $S^{(1)}(z,\vxw)$ denote the partial derivative of  $ S(z,\vxw)$ 
with respect to $z$.
Furthermore, we define
\bqan
\vD&=& a_1\E\big\{\vxw \vxw^T f(0|\vxw )\E(Y_1^{*2}+Y_0^{*2}|z=0,\ \vxw )\big\}, \label{vD}\\
\vQ&=&a_2\E\big\{\vxw \vxw^T f(0|\vxw )S^{(1)}(0,\vxw )\big\}, \label{vQ}
\eqan
where $f(z|\vxw )$ denotes the conditional probability density function of $z$ given $\vxw $,
$a_1 =2 \int \{K'(\nu )\}^2 d\nu$, and $a_2 = \int \nu K''(\nu) d\nu$,
with $K'(\cdot )$ and 
$K''(\cdot)$ denoting the first- and second-derivative of $K(\cdot)$, respectively. 
Note that $\vD$ and $\vQ$ both depend on unknown functions, e.g., $f(z|\widetilde{\vx})$, and are complex to approximate analytically. This 
	motivates us to consider a bootstrap approach for inference procedure.

\begin{theorem}
	\label{Th2}
Assume $K(\cdot)$ satisfies \ref{K1} -- \ref{K3} for some $b\geq 2$,
$h_n=o(n^{-1/(2b+1)})$ and  $n^{-1}h_n^{-4}=o(1)$. Then under \ref{A1} -- \ref{A5}, 

(1) $\sqrt{nh_n}(\vbetaw_n-\vbetaw_0) \ra N(\vnull, \vQ^{-1}\vD\vQ^{-1})$ in distribution as $n\ra \infty$.

(2) $\sqrt{n}\{V_n(\vbetah_n)-V(\vbeta_0)\} \ra  N(0, U)$ in distribution as $n\ra \infty$,
where $V_n(\cdot)$ is defined in (\ref{Vn}) and 
$U =\Var\{Y^*(d_{\vbeta_0})\}+\E\{(Y^*(d_{\vbeta_0})^2\}$.
\end{theorem}
\begin{remark}
Theorem~\ref{Th2} implies that $\vbetaw_n$ achieves a convergence rate arbitrarily close to $n^{-b/(2b+1)}$.
The cumulative distribution function of $N(0,1)$ satisfies these regularity conditions with $b=2$, 
and would produce a convergence rate arbitrarily close to $n^{-2/5}$. With a carefully designed $K(\cdot)$ function which satisfied  \ref{K1} -- \ref{K3} with $b$ sufficiently large, the convergence rate can be further improved. For example, 
$K(v) =\big[0.5 + \frac{105}{64}\{\frac{v}{5}-\frac{5}{3}(\frac{v}{5})^3 +\frac{7}{5}(\frac{v}{5})^5 - \frac{3}{7}(\frac{v}{5})^7\}\big]I( -5\leq v \leq 5)
+I(v>5)$ satisfies \ref{K1} -- \ref{K3} with $b=4$. 
This choice leads to a convergence rate of $n^{-4/9}$. 
This function first appeared in \citet{Horowitz}, which dealt with smoothing estimator in a different setting. Our setting and proofs are very different. Especially, our proofs substantially simplified the traditional methods for handling a smoothed objective function.
Example 2 in Section S7 of the supplementary material demonstrates that the performance of the smoothed estimator is not
 	sensitive to the choice of $K(\cdot)$ in finite samples. We would recommend the distribution function of N(0,1) as the default choice due to its simplicity, which we observe to have satisfactory performance in a variety of settings.  
\end{remark} 
\begin{remark} 
	The key components of the proofs are modern empirical process techniques. In particular, we introduce some recent empirical process results \citep{Sang2010, Mason} on VC classes of functions that involve smoothing parameters, which were originally developed for uniform asymptotics with data-driven bandwidth selection and have not been applied to the types of problems considered here. These new techniques lead to simpler proof and are of independent interest.
	Our technical derivation for this and other results in the paper employ recent techniques developed by \citet{Sang2010} and \citet{Mason} for VC classes of functions that involve smoothing parameters, see Appendix A. 
	Carefully handling function classes involving a smoothing parameter is nontrivial.
	The literature usually either impose a lower positive bound on $h$ to avoid the process to blow up or requires more involved computation on the entropy bound for such classes. In contrast, the new techniques are based on a geometric argument and avoid the usually intensive entropy computation. 
The asymptotic normality result in part (2) of the theorem is mostly due to the fact the estimated value function $V_n(\vbeta)$ is a sample average of functions that enjoy the
	Donsker property. Furthermore, the population value function $V(\vbeta)$ has gradient zero at the true value $\vbeta_0$. 
\end{remark}

\subsection{Justification for Resampling-based Inference} \label{Theory2}
Let $r_{1},...,r_{n}$ be a random sample from a distribution of a positive random variable with mean one and variance one. Assume  the random weights $r_{1},...,r_{n}$ are independent of the data.
Recall that
\bqa
\vbetah^*_n=\arg\max_{ \vbeta\in \bbB} \widetilde{M}^*_n(\vbeta)=\arg\max_{ \vbeta\in \bbB}\frac{2}{n}\sum_{i=1}^n r_{i}
 (2A_i-1) K\Big(\frac{\vx_i^T\vbeta}{h_n}\Big)Y_i.
\eqa
Hence, two different sources of randomness contribute to the distribution of $\vbetah^*_n$ in this setup: one due to the random data and the other due to the random weights.

We next provide a rigorous justification for the validity of the bootstrap procedures proposed in Section~\ref{dragon}. We establish that the bootstrap distribution asymptotically imitates the distribution of the original estimator. 
Let $r=\{r_1, \ldots, r_n\}$ be the collection of the random bootstrap weights and $w=\{W_1, \ldots, W_n\}$ be the random sample of observations, where $W_i=(\vx_i, A_i, Y_i)$.  

Given a sequence of random variables $R_n$, $n=1, \ldots, n$, we write $R_n=\opr$ if for any $\epsilon>0, \delta>0$,  we have $P_{w}(P_{r|w}(|R_n|>\epsilon)>\delta)\ra 0$ as $n\ra \infty$. In the bootstrap literature,  $R_n$ is said to converge to zero in probability, conditional on the data.  

\begin{theorem}
	\label{Th3}
	Under \ref{A1} -- \ref{A3}, \ref{A6} and assume $K(\cdot)$ satisfies \ref{K1}, then  
	
		(1) $\vbetah^*_n = \vbetah_n +\opr$;
		
		(2) $\sqrt{n}\big\{V_n^*( \vbetah_n)-V_n( \vbetah_n)\big\} =  N(0, U)+\opr$ $ \pxp$.
\end{theorem}

Part (2) of Theorem \ref{Th3} suggests that we can use the perturbed value function defined in (\ref{Vnstar}) with the plugged-in estimator $\vbetah_n$ to estimate the asymptotic variance of the estimated optimal value in Theorem~\ref{Th2}.
This establishes the asymptotic validity of the confidence interval in (\ref{b2}), which allows for inference for the value function. The validity of the confidence interval in (\ref{b1}) for $\vbeta_0$ is ensured by Theorem \ref{Th4} below.    
\begin{theorem}
	\label{Th4}
	Assume $K(\cdot)$ satisfies \ref{K1} -- \ref{K3} for some $b\geq 2$,
   $h_n=o(n^{-1/(2b+1)})$, and  $\log(n)=o(nh_n^4)$. Under  \ref{A1} - \ref{A6}, 
$	\sqrt{nh_n}(\vbetaw_n^*-\vbetaw_n)= N(\vnull, \vQ^{-1}\vD\vQ^{-1})+\opr.$
\end{theorem}


\begin{remark}
The proofs of Theorems \ref{Th3} and \ref{Th4} are given in Section S2 of the online supplementary material. We make use of the recent results in
 which allow for using an unconditional argument to derive conditional results. The use of the unconditional argument can be particularly convenient to combine with the Donsker class properties.  
\end{remark} 

To better understand the behavior of the proposed inference procedure, we also study the properties of the smoothed estimator and its bootstrapped version under a moving parameter or local asymptotic framework. See Section S4 of the online supplementary material.



\section{Simulation Results}\label{sec:simulate}
We generate random data from the model $Y= \exp(\vx^T\veta) + A\vx^T\vbeta + \epsilon$, where $\epsilon\sim N(0,1)$, $\vx = (x_0, x_1, x_2, x_3)^T=(x_0,\vxw^T)^T$, $x_0=1$ and $\vxw$ follows a 3-dimensional multivariate normal distribution with mean zero and identity covariance matrix. We set $\veta = (-1, -0.5,0.5,-0.5)^T$, and consider two settings for $\vbeta$. 
In setting 1, we have $\vbeta = (-2, -2, 2, 2)^T$; while in setting 2 we have $\vbeta = (-2, -2, 2, 0)^T$ with $x_3$ being an inactive variable for the optimal treatment regime. The optimal treatment regime is given by $\indicator(\vx^T\vbeta\leq 0)$.
As discussed in Section~\ref{setup}, for identifiability, we adopt the normalization $|\beta_1| = 1$, corresponding to the coefficient of the continuous covariate $x_1$. Under this normalization, the population parameter indexing the optimal treatment regime is $ \vbeta^{opt} = (\beta^{opt}_0, \beta^{opt}_1, \beta^{opt}_2,\beta^{opt}_3)=(-1, -1, 1, 1)$ in setting 1, and $(-1, -1, 1, 0)$ in setting 2. 
We consider 1000 simulation runs and three different sample sizes $n=300,\ 500,\ 1000$ in the simulation experiments.
The confidence intervals are constructed based on 500 bootstrap estimates for each simulation run. 
That is, for each simulation run, we generate 500 independent samples of size $n$ of positive random weights from a distribution with mean one and variance one and apply them to weight the original observations according to (\ref{boots}). 

We first study the finite sample performance of the smoothed robust estimator in Section~\ref{dragon}.
The smoothed robust estimator is computed using the proximal algorithm in Section \ref{algorithm}, where we choose $K(\cdot)$ to be the cumulative distribution function of standard normal distribution and set $h_n=0.9n^{-0.2} \min\{\mbox{ std} (\vx_i^T\vbeta),\mbox{ IQR}(\vx_i^T\vbeta)/1.34\}$, as suggested in \citet{Silverman1986}, where ``std'' denotes the standard deviation function, and ``IQR'' denotes the interquartile range.The initial estimator $\vbeta^0$ in the proximal algorithm is set as $(0,\cdots,0)^T$.
We compare the smoothed estimator with three alternative estimators. The first is the nonsmoothed estimator in (\ref{lion}), which was computed using the genetic algorithm, using the ``genoud'' function in R package ``rgenoud'' \citep{rgenoud}, as suggested in  \citet{zhang2012}. The second is the estimator based on the
hinge loss \citep{zhao2012estimating}, calculated using the function {\it owl} in the R package DTRlearn2 \citep{DTRlearn2}.  The 
third is the estimator using logistic loss, calculated using the function {\it glmnet} in the R package glmnet \citep{glmnet}. 
Table~\ref{table: Est} reports the bias and standard deviation of the estimate for the parameters indexing the optimal treatment regime, the match ratio (percentage of times the estimated optimal treatment regime matches the theoretically optimal treatment regime), and the bias and standard deviation of the estimated optimal value.

\begin{table}[!ht]
	\centering
	\def\~{\hphantom{0}}
	\caption{Monte Carlo estimates of 
		the bias and standard deviation of the estimate for the parameters indexing the optimal treatment regime, the match ratio (percentage of times the estimated optimal treatment regime matches the theoretically optimal treatment regime), and the bias and standard deviation of the estimated optimal value.
		}
	\label{table: Est}
	\resizebox{\textwidth}{!}{ 
	\begin{tabular} 
		{ccccccccc}
		\hline 
		$n$ & Method&$\beta^{opt}_0$&$\beta^{opt}_1$  &  $\beta^{opt}_2$  &  $\beta^{opt}_3$   &Match Ratio  & $V_n(\vbetah_n)$\\ 
		\hline
		\multicolumn{8}{c}{Setting 1}\vspace{0.5em}\\   
		\multirow{4}{*}{300}
		&Smooth & -0.05 (0.30) &  0 (0) &  0.01 (0.27) &  0.04 (0.31) & 99.35\% & -0.02 (0.17)   \\ 
		&Nonsmooth & -0.29 (1.45) &  0.00 (0.09) &  0.12 (1.21) &  0.24 (1.43) & 96.67\% &  0.06 (0.17)  \\ 
		&Hinge & -0.46 (0.41) &  0 (0) &  0.04 (0.27) & -0.04 (0.29) & 91.85\% & -0.05 (0.18)   \\ 
		&Logistic & -0.46 (0.47) &  0 (0) & 0.06 (0.42) &  0.26 (0.57) & 94.17\% & -0.02 (0.18)  \\\hline
		\multirow{4}{*}{500} &Smooth &  -0.01 (0.19) &  0 (0)&  0.01 (0.20) &  0.02 (0.22) & 99.73\% &  0.00 (0.13) \\ 
		&Nonsmooth & -0.15 (0.41) &  0 (0)&  0.06 (0.36) &  0.13 (0.42) & 98.19\% &  0.05 (0.13)\\ 
		&Hinge & -0.37 (0.30) &  0 (0) &  0.01 (0.18) & -0.06 (0.20) & 92.93\% & -0.03 (0.13)  \\ 
		&Logistic & -0.41 (0.29) &  0 (0)&  0.04 (0.30) &  0.23 (0.36) & 94.61\% & -0.01 (0.13)\\\hline
		\multirow{4}{*}{1000} 
		&Smooth &-0.01 (0.14) & 0 (0) &  0.00 (0.13) &  0.01 (0.15) & 99.88\% & -0.01 (0.09) \\ 
		&Nonsmooth & -0.07 (0.24) &  0 (0)&  0.02 (0.22) &  0.06 (0.25) & 99.04\% &  0.03 (0.09) \\ 
		&Hinge &  -0.36 (0.24) &   0 (0)&   0.01 (0.13) &  -0.07 (0.14) &  92.95\% &  -0.04 (0.09)  \\ 
		&Logistic & -0.38 (0.19) &  0 (0) &  0.02 (0.19) &  0.18 (0.23) & 94.61\% & -0.02 (0.09)  \\ 
		\hline 
		\multicolumn{8}{c}{Setting 2}\vspace{0.5em}	\\ 
		\multirow{4}{*}{300} &Smooth &0.04 (0.26) &  0 (0)  &  0.02 (0.24) &  0.02 (0.18) & 99.35\% & -0.01 (0.15) \\ 
		&Nonsmooth & -0.26 (0.76) &  0.00 (0.06) &  0.11 (0.71) &  0.11 (0.37) & 95.78\% &  0.07 (0.15) \\ 
		&Hinge & -3.33 (79.42) &   0 (0)  &  0.01 (0.22) & -0.09 (0.16) & 76.19\% & -0.06 (0.16)  \\ 
		&Logistic & -0.67 (5.13) &  0.00 (0.06) &  0.18 (3.33) &  0.23 (2.96) & 90.20\% & -0.02 (0.16)\\\hline
		\multirow{4}{*}{500} & Smooth& 0.02 (0.19) &  0 (0)&  0.02 (0.18) &  0.00 (0.13) & 99.65\% & -0.01 (0.11) \\ 
		&Nonsmooth & -0.16 (0.52) &  0 (0) &  0.06 (0.42) &  0.06 (0.24) & 97.37\% &  0.05 (0.11)  \\ 
		&Hinge & -0.64 (1.11) &  0 (0)&  0.02 (0.16) & -0.10 (0.12) & 88.59\% & -0.07 (0.12)  \\ 
		&Logistic & -0.43 (0.29) &  0 (0) &  0.03 (0.30) &  0.12 (0.20) & 92.08\% & -0.03 (0.12)\\\hline
		\multirow{4}{*}{ 1000}  & Smooth&  -0.01 (0.14) & 0 (0) &  0.01 (0.13) &  0.00 (0.09) & 99.79\% & -0.01 (0.08)   \\ 
		&Nonsmooth & -0.08 (0.21) &  0 (0)  &  0.03 (0.22) &  0.04 (0.17) & 98.55\% &  0.03 (0.08) \\ 
		&Hinge &  -0.56 (0.24) &    0 (0) &   0.01 (0.12) &  -0.10 (0.08) &  89.69\% &  -0.06 (0.09) \\ 
		&Logistic & -0.43 (0.20) &  0 (0) &  0.03 (0.20) &  0.11 (0.15) & 92.13\% & -0.03 (0.09) \\
		\hline
	\end{tabular}}  \vskip 18pt
\end{table}

The results in Table~\ref{table: Est} demonstrates that the smoothed robust estimate has smaller bias and substantially smaller standard deviation comparing with theother three estimators, particular for the smaller sample size setting. It also leads to higher match ratio. Estimators using hinge loss and logistic loss are even not consistent when the sample size increases.
For $n=300$, we observe that in one or two of the 100 simulation runs
the non-smooth estimator converges to the negative of the true value of $\beta_1^{opt}$
(i.e., the algorithm converges to 1 when the true value is -1), which causes the non-zero variance.
This is probably due to the fact nonsmooth estimation is less stable when the sample size is relatively small.
In addition, the expected value functions with the true parameter $\vbeta^{opt}$ and random policy are simulated via Monte Carlo simulation with $10^7$ replicates; for Setting 1, the optimal value turns out to be 1.14, and the value function with random policy is -0.47; and for Setting 2, the true optimal value is 0.93, and the value function with random policy is -0.29. 
When taking the computation time into consideration, the nonsmoothed estimator requires about 4 seconds for each run, while  the smoothed estimator only needs 0.002 seconds. This suggests a substantial reduction in computational costs.

We next investigate the bootstrap confidence interval  in Section \ref{dragon}.  We construct 95\% bootstrap confidence intervals for the parameters indexing the optimal treatment regime.
Table~\ref{table: Bootstrap} summarizes the empirical coverage probabilities and average interval lengths. 
We observe that the 
coverage probabilities are above 92.2\% for sample sizes 500 and 1000, and above 91\% for sample size 300. 
Despite the slight under coverage, the lengths of the confidence intervals are reasonable.
As sample size increases, the length of the confidence interval decreases significantly. 
Accurate finite-sample coverage is harder to achieve due to the model-free, nonparametric nature of our approach. 
See similar observations in simulations focusing on non-regularity settings for dynamic treatment regimes, for instance, \cite{laber2010} and \citet{chakraborty2013}. As for computation time, on average one bootstrap run takes less than 0.2 seconds.
 
\begin{table}[!ht] 
	\centering  
	\def~{\hphantom{0}} 
	\caption{Empirical coverage probabilities and average interval lengths of the 95\% bootstrap confidence intervals for $\vbeta^{opt}$}
	\label{table: Bootstrap} 
		\begin{tabular}{cccccc}
			\hline
$n$ & &$\beta^{opt}_0$&$\beta^{opt}_1$  &  $\beta^{opt}_2$  &  $\beta^{opt}_3$  \\
[0.5em]\hline
\multicolumn{6}{c}{Setting 1} \\
[0.5em] 
 \multirow{2}{*}{300}  &\CR&92.6\% & 100\%  & 93.2\% & 91.0\% \\
 & \AL&1.36 &0 &  1.26&1.38 \\
 \multirow{2}{*}{500}  & \CR&92.2\% & 100\%  & 93.0\% & 92.6\%  \\
 &  \AL& 0.81&0&0.79&0.84 \\
 \multirow{2}{*}{1000}  & \CR&92.6\% & 100\%  & 94.0\% & 93.4\% \\
 & \AL&0.54&0&0.53&0.56\\
 [0.5em]\hline
 \multicolumn{6}{c}{Setting 2} \\
 [0.5em] 
 \multirow{2}{*}{300}  &\CR &93.4\% & 100\%  & 92.6\% & 95.8\%\\
 &  \AL&1.12&0&1.01&0.71  \\
 \multirow{2}{*}{500}  & \CR&94.2\% & 100\%  & 93.8\% & 94.6\%\\
 & \AL &0.75&0&0.72&0.51 \\
 \multirow{2}{*}{1000}  &\CR &94.0\% & 100\%  & 93.0\% & 95.4\%\\
 & \AL&0.50&0&0.48&0.35 \\\hline
\end{tabular}\vskip 18pt
\end{table} 

Finally, we explore several nonregular settings, where the optimal treatment regimes may be nonunique, motivated by \cite{laber2010}. 
In these cases, the parameter indexing the optimal treatment regime is not uniquely identifiable but inference for the optimal value may still be feasible. We focus here on the bootstrap confidence interval for the optimal value.  In setting 3,  the same data generative model as before is used with $\vbeta = (1, 2, 0.02, 0)^T$. For setting 4 and 5, $\vbeta = (-1, 1, 0, 0)^T$, however, the first random covariate $x_1$ is generated from the discrete uniform distribution on the set $\{-1,0,1,2\}$ and $\{1,2\}$, respectively, instead of the standard normal distribution. 
For completeness, the bootstrap confidence intervals for the optimal value in setting 1 and setting 2 are also studied.

Let $p$ denote the probability of generating a covariate vector $\vx$ such that $\vx^T\vbeta=0$. This is a useful measure of the nonregularity of  the model \citep{laber2010}. According to this measurement, setting 1 -- 3 are regular (R) cases with $p=0$; while setting 4 and 5 are nonregular (NR) with $p=0.25$ for setting 4 and $p=0.5$ for setting 5.


Table~\ref{table: value} summarizes the empirical coverage rate and average length for the 95\% bootstrap confidence intervals for the optimal value functions. The results demonstrate that the bootstrap confidence intervals for the optimal value have desirable coverage rates with reasonable interval lengths, even in the nonregular cases. For comparison, we also report the percentage of times these bootstrap confidence would cover the value function from a random policy. The percentage is really low, which implies that the proposed method performs much better than random assignment even in the nonregular cases. 

\begin{table}[!ht]
	\centering
	\def\~{\hphantom{0}}
		\caption{Empirical coverage probabilities and average interval lengths of the 95\% confidence intervals for $V(\vbeta^{opt})$}
		\label{table: value}
		\begin{tabular}{ccccccc}
			\hline 
			&Setting&1&2&3&4&5\\ 
			$n$ &Type& R & R & R & NR & NR\\\hline
			\multirow{3}{*}{300} &  \CR&93.0\%& 92.6\%& 96.4\%& 97.2\%& 95.4\%\\
			&\AL&0.67&  0.61& 0.78& 0.40& 0.41\\
			&  CR for random policy&0\%& 0\%& 0\%& 0\%& 31.2\%\\ \hline
			\multirow{3}{*}{500} &  \CR&93.8\%& 94.0\%& 96.0\%& 95.2\%& 94.4\%\\
			&\AL&0.52&  0.47& 0.62& 0.31& 0.31\\ 
			&  CR for random policy&0\%& 0\%& 0\%& 0\%& 12.4\%\\ \hline
			\multirow{3}{*}{1000} &  \CR& 93.6\%& 95.4\%& 97.0\%& 96.0\%& 96.0\%\\
			&\AL&0.37&  0.33& 0.43&0.22& 0.22\\
			&  CR for random policy&0\%& 0\%& 0\%& 0\%& 0.8\%\\\hline
			\hline
		\end{tabular} 
\end{table} 

\section{A Real Data Example}\label{sec:realdata}
We analyze a clinical data set from the Childhood Adenotonsillectomy Trial (CHAT). This is a randomized study designed to test whether early adenotonsillectomy (eAT, denoted as treatment 1) is helpful to improve neurocognitive functioning, behavior and quality of life for children with mild to moderate obstructive sleep apnea, compared with watchful waiting plus supportive care (WWSC, denoted as treatment 0), see \citet{sleep1}. In this trial, 464 children with mild to moderate obstructive sleep apnea syndrome, ages 5 to 9.9 years, were randomly assigned to eAT and WWSC. Some biochemical and neurocognitive test results were recorded before the treatment and seven months after the treatment.

We consider the baseline Apnea-Hypopnea Index (AHI),  with a natural log-transformation as recommended by \citet{sleep1},  as an explanatory variable. AHI is the number of apneas or hypopneas recorded during the study per hour of sleep. It is an important measurement of the quality of sleep and is commonly used by doctors to classify the severity of sleep apnea. \citet{sleep1} suggested that  black children tend to experience different improvements with eAT comparing with children from other races. We hence include race (binary, 1=African American, 0 for others) as another covariate. For the outcome variable, to balance the benefits and adverse effects from eAT, we adopt a composite score. 
The composite score uses the ratio of the follow-up AHI and baseline AHI  (both with natural log-transformations) as an effective measure of benefit. On the other hand, it takes into account the adverse events documented according to the CHAT study manual of procedures as penalty. 

We estimate the optimal treatment regime in the class of treatment regimes $\bbD = \{I(\beta_0+\beta_1 \mbox{AHI}+\beta_2 \mbox{race}>0):  |\beta_1|=1\}$. The kernel function $K(\cdot)$ and the bandwidth selection are the same as in Section~\ref{sec:simulate}.  The smoothed estimator for the baseline AHI is normalized to 1, the race is 0.56, with $(0.34, 0.97)$ as the 95\% bootstrap confidence interval, and the intercept is 0.39, with confidence interval $(0.22, 0.65)$. The confidence intervals suggest that the coefficients are all significantly different from 0.
The analysis suggests that it is reasonable to assign WWSC to those children with milder symptoms (lower AHI). It also suggests that black children display more improvement in the AHI scale with eAT.  The results are consistent with those observed empirically in  
\citet{sleep2},  \citet{sleep1} and \citet{sleep3}.  
The average outcome with randomized treatment is 0.288. The estimated average outcome corresponding to the estimated optimal treatment regime is 0.063, with a 95\% bootstrap confidence interval  $(-0.126, 0.260)$.  This suggested a significant reduction of the composite outcome score when applying the optimal treatment regime.To compare with the smoothed estimator, we also calculate the nonsmoothed estimator, whose coefficients are 1 for baseline AHI, -0.19 for the race, and -0.40 for the intercept. Its estimated optimal value is -0.034. The nonsmoothed estimators are significantly different from the smoothed ones. In Example 4 of Section S7 in the supplementary, we demonstrate based on five-fold cross-validation that for this real data example, the nonsmoothed estimator is quite unstable.

%

\section{Discussions}\label{sec:diss}

\subsection{Extension to other settings}
The method we propose can be extended to observational studies
using the inverse probability weighting approach.
Assume the propensity score $\pi(x)=P(A=1|\vx)$ can be modeled as $\pi(\vx, \vxi)$ where $\vxi$ is a finite-dimensional parameter
(e.g., via logistic regression). Let $\widehat{\vxi}$ be an estimate of $\xi$.
Under the commonly adopted assumption of no unmeasured confounding, a smoothed robust estimator for $\vbeta_0$ can be constructed as 
 \bqan\label{obs}
\arg\max_{\vbeta\in \bbB}n^{-1}\sum_{i=1}^n
\frac{\big[A_iK\big(\frac{\vx_i^T\vbeta}{h_n}\big)+(1-A_i)\big\{1-K\big(\frac{\vx_i^T\vbeta}{h_n}\big)\big\}\big]Y_i}
{A_i\pi(\vx, \widehat{\vxi}) +(1-A_i)(1-\pi(\vx, \widehat{\vxi}))}.
\eqan
 Example 3 in Section S7 of the supplementary material confirms that this smoothed estimator 
provides accurate estimation for the optimal treatment regime 
when the propensity score model is correctly specified. 
The estimator in (\ref{obs}) can also be extended to be doubly robust similarly as in \citet{zhang2012}. 
Due to the presence of nuisance parameter, the theory of asymptotic normality and inference is more technically involved.
This will be a future research topic.

It is worth pointing out that our method is applicable to binary response, as binary random variable is sub-Gaussian after centering. 
Example 1 in Section S7 of the supplementary materiel demonstrates that our estimation and inference procedures
work effectively for binary responses. For survival outcome under random censoring, our method can be extended
to obtain a robust procedure for estimating the optimal treatment regime maximizing the restricted mean survival time, similarly as in \citet{zhao2015doubly}. 
Let $\widetilde{T}$ denote the survival time. Let $T=\min\{\widetilde{T}, \tau\}$ be the outcome of interest, where $\tau$ is the time till the end of the study.
Let $C$ denote the censoring time and $\Delta = I(T<C)$ be the censoring indicator.
We observe $Y=\min\{T, C\}$. Based on the observed data $\{Y_i, \vx_i, \Delta_i, A_i\}$, $i=1, \ldots, n$ from a randomized trial,
the smoothed estimator can be constructed as
\begin{align*}
\arg\max_{\vbeta\in \bbB}\frac{2}{n}\sum_{i=1}^n
\frac{\big[A_iK\big(\frac{\vx_i^T\vbeta}{h_n}\big)+(1-A_i)\big\{1-K\big(\frac{\vx_i^T\vbeta}{h_n}\big)\big\}\big]}
{\widehat{G}_C(Y_i|\vx,A)}\Delta_iY_i,
\end{align*}
where $G_C(t|\vx,A)=P(C>t|\vx,A)$ is the conditional survival function of the censoring time $C$ given $(\vX,A)$, 
and $\widehat{G}_C(\cdot|\vx,A)$ is an estimator of $G_C(\cdot|\vx,A)$.



\subsection{On the identifiability condition}
The asymptotic normality results can be established under alternative identifiability constraint
		such as the requirement that the $L_1/L_2$ norm of $\vbeta$ is 1, or identifiability of $\vbeta$ up to a scale. 
		However, this usually leads to more technically involved 
		proof as $\vbeta$ is constrained to be the boundary point of a unit sphere and $V(\vbeta)$ does not have a derivative at $\vbeta$.
		This issue was often ignored in the theory development in many existing literature, which only adjust for the constraint 
		in an ad-hoc way in the numerical implementation. See \citet{zhu2006}
for more discussions in an index model setting and a careful delete-one-component method to handle this rigorously.

For identifiability, we assume that there exists a covariate 
	whose conditional distribution given the other covariates is absolutely continuous. This is a common assumption for index model and is satisfied in many real applications. 
	In practice, domain experts may help suggest such a candidate continuous covariate and the statisticians can run confirmatory
	analysis (e.g., comparing the conditional treatment effect conditional on this covariate) to verify if this is a viable choice.
	In the case when all relevant covarites are discrete (e.g, gender, race), the problem reduces to comparing a finite number of decision rules and the main target of inference is arguably the optimal value. Our simulation settings 4 \& 5 only include discrete variables in the optimal regime. 
	The simulation results in Table 3 show that our proposed bootstrap 
	confidence interval still provides reasonable empirical coverage probability for the optimal value in discrete cases.

\subsection{Non-regular settings}
The optimal treatment regime may not be unique if there exists a subpopulation who responds similarly to the two treatment options. 
In such a setting, the complexity of nonregularity arises, see the discussions in \citet{Robins2004}, \citet{moodie2010}, \cite{laber2010},  
\citet{Song2015}, and  \citet{Luedtke2016}. Uniform inference under nonregularity or exceptional laws is a challenging problem. 

Although our theory does not apply to this scenario, our simulation results show that our bootstrap confidence interval 
for the optimal value function displays a fair degree of robustness in the two examples where nonregularity occurs. 
As an example, in simulation setting 5, if $x_1=1$, then the subject responds the same to the two treatment options; while if $x_1=2$, the subject benefits from treatment 1.  There are four decision rules of interest for this example. The optimal treatment rule is nonunique as one may assign either treatment 0 (say no treatment or a standard, less expensive treatment) or treatment 1 to those subjects with $x_1=1$. 
A relative simple approach to breaking the nonuniqueness is to introduce a secondary criterion.
For example, one may argue that under the principle of avoiding over-treatment, there exists a unique optimal decision rule of interest, in this case $I(x_1=2)$, which would not assign treatment 1 when ambiguity exists in order to reduce costs and avoid potential risks. 
Based on the sample, this unique optimal treatment regime can be consistently estimated by selecting the decision rule that maximizes the sample average treatment effect while treating the smallest proportion of the population. 
 
There are additional inference targets that have rarely been discussed in the literature, that is, inference about the linear combination in the rule $\vx^T\vbeta$ or about the rule itself $I(\vx^T\vbeta>0)$. These two quantities are of interest in clinical practice as they indicate how much confidence we can put on the prescribed optimal decision. We are currently studying these inference problems and will report the results in a future article.

 \section*{Acknowledgments}
 The authors thanks the co-Editor, the AE and two anonymous referees for their constructive comments which have helped us significantly
 improve the paper. 
 The authors acknowledge financial support from NSF DMS-1712706, NSF OAC-1940160 and NSF FRGMS-1952373. 

  \bibliographystyle{plain} 
\bibliography{Lreference1}   
\appendix

\section{Regularity Conditions and Useful Lemmas} 

We first state some regularity conditions, where \ref{K1}--\ref{K3}  are assumptions imposed on $K(\cdot)$, while
\ref{A1}--\ref{A6} are assumptions imposed on the data. 
\begin{enumerate}[label=(K\arabic*),wide=0pt]
\item \label{K1} $K(\cdot)$ is twice differentiable, $K(\cdot)$, $K'(\cdot)$ and $K''(\cdot)$ all bounded variation on the real line.  Furthermore,
$\lim\limits_{\nu \rightarrow -\infty} K(\nu)= 0$, $\lim\limits_{\nu \rightarrow \infty} K(\nu)= 1$; 
$\int \{K'(\nu)\}^2 d\nu$ and $\int \{K''(\nu)\}^2 d\nu$ are both finite. 
\item \label{K2} For some integer $b \geq 2$, and any $1\leq i \leq b$,
$\int |\nu^iK'(\nu)| d\nu< \infty$;  $\int_{-\infty}^{\infty} \nu^iK'(\nu) d\nu=0$ for $1\leq i\leq b-1$ and $\int_{-\infty}^{\infty} \nu^bK'(\nu) d\nu=d\neq 0$. 
\item \label{K3} For any integer $i$ between $0$ and $b$, any $\eta > 0$, and any sequence $\{h_n\}$ converging to 0,  $ \lim\limits_{n\ra\infty} h_n^{i-b}\int_{|h_nv|>\eta}  |\nu^iK'(\nu)| d\nu = 0$, and	$\lim\limits_{n\ra\infty} h_n^{-1}\int_{|h_n\nu|>\eta}  |K''(\nu)| d\nu = 0$. 
\end{enumerate} 
\begin{enumerate}[label=(A\arabic*),wide=0pt]	
\item \label{A1} $\mu(a,\vx)$ is bounded for almost all $\vx$, and $a=0,1$; $Y_a^*-\mu(a,\vx)$, $a=0, 1$, has a sub-Gaussian distribution for almost every $\vx$. 
\item \label{A2} The support of the distribution of $\vx$ is not contained in any proper linear subspace of $\bbR^p$.  For almost every $\vxw $, the distribution of $x_1$ conditional on $\vxw$ has everywhere a positive density. 	  The components of $\vxw $ are bounded by $M_x$. 
\item \label{A3} Let $ S(z,\vxw) = \E\{Y^*_1-Y^*_0|z,\vxw\}$, where $z=\vx^T\vbeta_0$.   For almost every $\vxw $, $S(0,\vxw )=0$. 	And for every $\epsilon>0$, 	$\sup_{||\vbeta-\vbeta_0|| > \epsilon}\E\{\indicator(x^T\vbeta>0)S(z,\vxw ) f(z|\vxw )\} < \E\{\indicator(x^T\vbeta_0>0)S(z,\vxw ) f(z|\vxw )\}$.  
\item  \label{A4} Given any integer $0\leq i\leq b-1$, for  all $z$ in a neighborhood of $0$, $f^{(i)}(z|\vxw )$ is a continuous function of $z$  and satisfies $|f^{(i)}(z|\vxw )| <M_f$ for almost every $\vxw $, where $M_f>0$ is a constant.  
\item  \label{A5}Let $S^{(i)}(0,\vxw)$, $i=0, 1,\ldots, b$, denote the $i$th partial derivative of  $ S(z,\vxw)$  with respect to $z$. For $0\leq i\leq b$, for all $z$ in a neighborhood of $0$, $S^{(i)}(z,\vxw )$ is a continuous function of $z$ and satisfies $|S^{(i)}(z,\vxw )| <M_S$ for almost every $\vxw $, where $M_s>0$ is a constant.  The matrices $\E\{\vxw \vxw^T f(0|\vxw )S^{(1)}(0,\vxw )\}$  and $-\E\{\vxw \vxw^T (\vxw^T\vbetaw_0) f(0|\vxw )S^{(1)}(0,\vxw )\}$ are negative definite.  
\item \label{A6}  The random weights $r_{1},...,r_{n}$ form a random sample from a distribution of a  positive random variable with mean one and variance one. Assume that $r_i-\E(r_i)$ has a sub-Gaussian distribution, $i=1, \ldots, n$.
\end{enumerate}  
\begin{remark} 
The bounded variation assumption on $K(\cdot)$, $K'(\cdot)$ and $K''(\cdot)$ are relatively weak (Chapter 6, \citet{Apostol1974}). This and other assumptions in \ref{K1}-\ref{K2} are satisfied if $K(\cdot)$ is taken to be the distribution function of standard normal distribution ($b=2)$ or the function in Remark 1 ($b=4$).
However, $K(\cdot)$ is not required to be a cumulative distribution function.
The bounded variation assumption implies that $K(\cdot)$, $|K'(\cdot)|$ and $|K''(\cdot)|$ are uniformly bounded.
Our assumptions on the data are also relatively mild.
Condition \ref{A1} imposes mild assumption on the tail distribution of $Y_a^*-\mu(a,\vx)$, $a=0, 1$, and allows for both normal distribution and many other nonnormal distributions.
Condition \ref{A3} is a margin type condition to ensure identification of $\vbeta_0$.
\end{remark}

Let 
\begin{align*}
\mathcal{G} =&\big\{A\indicator(\vx^T\vbeta>0)Y + (1-A)\indicator(\vx^T\vbeta\leq 0)Y: \vbeta\in\bbB\big\},\\
\mathcal{G}^*=&\Big\{(r-1)\{A\indicator(\vx^T\vbeta>0)+ (1-A)\indicator(\vx^T\vbeta\leq 0)\}Y: \vbeta\in\bbB\Big\}.
\end{align*} 
 It is easy to see $\mathcal{G}$ and  $\mathcal{G}^*$ are both Donsker classes of functions. Next, we state a useful lemma concerning the Donsker properties of several other classes of functions that involve a smoothing parameter, as well as four technical lemmas that are useful for the proof of the main theorems and are proved based on the Donsker properties using empirical processes techniques. Their proofs can be found in the online supplementary material.

\setcounter{lemma}{0}
\renewcommand{\thelemma}{A\arabic{lemma}} 
\begin{lemma}
\label{Moon} 
Under \ref{K1}, \ref{A1}-\ref{A3}, the following six classes of functions are Donsker classes. 
\begin{align*}
\mathcal{F} =&\big\{(2A-1)K\big(\frac{\vx^T\vbeta}{h}\big)Y: \vbeta\in\bbB, h\in (0,1]\big\},\\ \mathcal{F}^*=&\big\{r(2A-1)K\big(\frac{\vx^T\vbeta}{h}\big)Y: \vbeta\in\bbB, h\in (0,1]\big\},\\
\mathcal{H} =&\big\{(2A-1)K'\big(\frac{z+\vpsi^T\vxw}{h}\big)\vxw Y: \vpsi\in\Psi, h\in (0,1]\big\}, \\
\mathcal{H}^*=& \big\{r(2A-1)K'\big(\frac{\vx^T\vbeta}{h}\big) \vxw Y:\vbeta\in\bbB, h\in(0,1] \big\},\\
\mathcal{Q} =& \big\{(2A-1)K''\big(\frac{\vx^T\vbeta}{h}\big) \vxw\vxw^T Y:\vbeta\in\bbB, h\in(0,1] \big\}, \\
\mathcal{Q}^*=&\big \{r(2A-1)K''\big(\frac{\vx^T\vbeta}{h}\big) \vxw\vxw^T Y:\vbeta\in\bbB, h\in(0,1]\big\}, 
\end{align*}
 
where $\Psi =\{\vpsi: \vpsi\in\bbR^{p-1}, ||\vpsi||\leq \frac{\eta}{2\sqrt{p-1}M_x}\}$, with $||\cdot||$ denoting the $l_2$ norm.
	\end{lemma}

\begin{lemma}\label{Lemma5} 
    Let $G_i(\vx_i,\vbeta,h_n)=(2A_i-1)K\big(\frac{\vx_i^T\vbeta}{h_n}\big)Y_i - \E \big\{(2A_i-1)\indicator(\vx_i^T\vbeta>0)Y_i\big\} $. 
	Under Assumptions \ref{A1}-\ref{A3} and \ref{K1},
 $\sup\limits_{\vbeta\in\bbB}\big|n^{-1}\sum_{i=1}^{n}G_i(\vx_i,\vbeta,h_n)\big|\rap 0$.
\end{lemma}

\begin{lemma}
	\label{L3}
	For any $\vtheta \in \bbR^{p-1}$, let
$\vR_n(\vtheta) = \frac{2}{nh_n^2}\sum_{i=1}^n(2A_i-1)K'\big(\frac{z_i}{ h_n}+  \vtheta^T \vxw _i\big) \vxw _iY_i$.
let $\eta >0$ be such that $S^{(1)}(z,\vxw )$, $S^{(2)}(z,\vxw )$, and $f^{(1)}(z|\vxw )$ exist and are uniformly bounded for almost 
every $\vxw $ if $|z| \leq \eta$.  Define 
$\Theta_n = \big\{\vtheta : \vtheta\in \bbR^{p-1}, h_n||\vtheta|| \leq \frac{\eta}{2\sqrt{p-1}M_x} \big\}$.
Assume the conditions of Theorem~\ref{Th2} are satisfied, then
	(1) $\sup\limits_{\vtheta\in\Theta_n}||\vR_n(\vtheta)-\E \vR_n(\vtheta)||\rap 0 $.
	(2) There are finite numbers $\alpha_1$ and $\alpha_2$ such that for all $\vtheta\in\Theta_n$,
	$||\E \vR_n(\vtheta) - \vQ\vtheta|| \leq o(1)+\alpha_1h_n||\vtheta|| +\alpha_2h_n||\vtheta||^2$
	uniformly over  $\vtheta\in\Theta_n$.
\end{lemma}

\begin{lemma}
	\label{Lemma10}
	Define 
	$G^*_i(\vx_i,\vbeta,h_n)=(2A_i-1)r_iK\big(\frac{\vx_i^T\vbeta}{h_n}\big)Y_i - \E \big\{r_i(2A_i-1)\indicator(\vx_i^T\vbeta>0)Y_i\big\}$ .
	Under Assumptions \ref{A1}-\ref{A3} and \ref{K1},
 $\sup\limits_{\vbeta\in\bbB}\big|n^{-1}\sum_{i=1}^{n}G^*_i(\vx_i,\vbeta,h_n)\big| = \oprw$,
 where $\oprw$ denotes a random sequence that converges to zero in probability with respect to the joint distribution of $(r, w)$.
\end{lemma}

\begin{lemma}
	\label{L9}
	Assume the conditions of Theorem~\ref{Th4} are satisfied, then
$(nh_n)^{1/2}\big\{T_n^*(\vbetah_n; h_n)-T_n^*(\vbeta_0; h_n)\big\}=o_{p_r}(1)$,
	where $T_n^*(\vbeta,h_n)$ is defined as follows:
	\begin{align*}
	\vT_n^*(\vbeta; h_n) &= \frac{\partial \widetilde{M}^*_n(\vbeta, h_n)}{\partial \vbetaw} =  \frac{2}{n} \sum_{i=1}^n r_i(2A_i -1) K'\big(\frac{\vx_i^T\vbeta}{ h_n} \big) \frac{\vxw_i}{h_n} Y_i. 
	\end{align*}
\end{lemma}

 \section*{Supplementary material}
The supplementary material is constructed as follows. 	Section~\ref{sample_value} presents and proves several useful lemmas.	Section~\ref{Proofs} proves Theorem \ref{Th1}--\ref{Th4} in the main paper. The additional technical lemmas appeared in Section~\ref{Proofs} are proved in Section~\ref{Smoothproof}. In Section~\ref{moving},  we study the properties of the smoothed estimator and its bootstrapped version under a moving parameter or local asymptotic framework. Proofs of Theorem~\ref{Th5} and Theorem~\ref{Th6} in Section~\ref{moving} are given in Section~\ref{Movingproof}.  Section~\ref{Algo} presents the pseudo codes for the proximal algorithm in Section~\ref{algorithm} of the main paper. Section~\ref{numeric} presents some additional numerical results.

\setcounter{section}{0} 
\setcounter{equation}{0} 
\setcounter{lemma}{0} 
\renewcommand{\thesection}{S\arabic{section}}  
\renewcommand{\theequation}{S\arabic{equation}} 
\renewcommand{\thelemma}{\arabic{lemma}}

\section{Some Useful Lemmas}\label{sample_value}

\setcounter{lemma}{0}
\renewcommand{\thelemma}{S\arabic{lemma}} 

\begin{lemma}
	\label{value_missing}
	$\E\big[\big\{\indicator(A=1)\indicator(\vx^T\vbeta>0)  +\indicator(A=0)\indicator(\vx^T\vbeta\leq 0) \big\}Y\big] = \frac{1}{2}V(\vbeta)$.
\end{lemma}
\begin{newproof}{ Lemma~\ref{value_missing}}
	By the iterative expectation formula,
	\begin{align*}
	&\E\big[\big\{\indicator(A=1)\indicator(\vx^T\vbeta>0)  +\indicator(A=0)\indicator(\vx^T\vbeta\leq 0) \big\}Y\big] \\
	=& \E_{A,\vx}\big[\big\{\indicator(A=1)\indicator(\vx^T\vbeta>0)  +\indicator(A=0)\indicator(\vx^T\vbeta\leq 0) \big\} E(Y|A,\vx)\big]\\
	=&\E_{A,\vx } \big\{\indicator(A=1)\indicator(\vx^T\vbeta>0)\mu(1,\vx) +\indicator(A=0)\indicator(\vx^T\vbeta\leq 0)\mu(0,\vx)\big\}\\
	=&\frac{1}{2}\E_{\vx}\big\{\indicator(\vx^T\vbeta>0)\mu(1,\vx)+\indicator(\vx^T\vbeta\leq 0)\mu(0,\vx)\big\} = \frac{1}{2}V(\vbeta).
	\end{align*} 
\end{newproof}

Let \vspace{-2em}
\begin{align*}
\mathcal{G} =&\big\{A\indicator(\vx^T\vbeta>0)Y + (1-A)\indicator(\vx^T\vbeta\leq 0)Y: \vbeta\in\bbB\big\},\\
\mathcal{G}^*=&\Big\{(r-1)\{A\indicator(\vx^T\vbeta>0)+ (1-A)\indicator(\vx^T\vbeta\leq 0)\}Y: \vbeta\in\bbB\Big\}.
\end{align*} 
It is easy to see $\mathcal{G}$ and  $\mathcal{G}^*$ are both Donsker classes of functions. Next, we state a useful lemma concerning the Donsker properties of several other classes of functions that involve a smoothing parameter, as well as four technical lemmas that are useful for the proof of the main theorems and are proved based on the Donsker properties using empirical processes techniques. Their proofs can be found in the following.

\setcounter{lemma}{0}
\renewcommand{\thelemma}{A\arabic{lemma}} 

\begin{lemma}
	Under \ref{K1}, \ref{A1}-\ref{A3}, the following six classes of functions are Donsker classes. 
	\begin{align*}
	\mathcal{F} =&\big\{(2A-1)K\big(\frac{\vx^T\vbeta}{h}\big)Y: \vbeta\in\bbB, h\in (0,1]\big\},\\ \mathcal{F}^*=&\big\{r(2A-1)K\big(\frac{\vx^T\vbeta}{h}\big)Y: \vbeta\in\bbB, h\in (0,1]\big\},\\
	\mathcal{H} =&\big\{(2A-1)K'\big(\frac{z+\vpsi^T\vxw}{h}\big)\vxw Y: \vpsi\in\Psi, h\in (0,1]\big\}, \\
	\mathcal{H}^*=& \big\{r(2A-1)K'\big(\frac{\vx^T\vbeta}{h}\big) \vxw Y:\vbeta\in\bbB, h\in(0,1] \big\},\\
	\mathcal{Q} =& \big\{(2A-1)K''\big(\frac{\vx^T\vbeta}{h}\big) \vxw\vxw^T Y:\vbeta\in\bbB, h\in(0,1] \big\}, \\
	\mathcal{Q}^*=&\big \{r(2A-1)K''\big(\frac{\vx^T\vbeta}{h}\big) \vxw\vxw^T Y:\vbeta\in\bbB, h\in(0,1]\big\}, 
	\end{align*}
	
	where $\Psi =\{\vpsi: \vpsi\in\bbR^{p-1}, ||\vpsi||\leq \frac{\eta}{2\sqrt{p-1}M_x}\}$, with $||\cdot||$ denoting the $l_2$ norm.
\end{lemma}

\begin{newproof}{ Lemma~\ref{Moon}}
	We give below the proof for $\mathcal{F}$. Proofs for the other classes of functions are similar. Since $K(\cdot)$ is continuous, and has bounded variation on  the real line, by Jordan's Theorem in Section 6.3 in \citet{Royden2010real}, there exist bounded, nondecreasing, right continuous functions 	$K_1$ and $K_2$ on $\bbR$ such that $K = K_1-K_2$.  Let 
	$\mathcal{F}_1=\big\{(2A-1)K_1\big(\frac{\vx^T\vbeta}{h}\big) Y: \vbeta\in\bbB, h\in (0,1]\big\}$,
	and
	$\mathcal{F}_2=\big\{(2A-1)K_2\big(\frac{\vx^T\vbeta}{h}\big) Y: \vbeta\in\bbB, h\in (0,1]\big\}$.
	Furthermore, let $\mathcal{F}_{10}=\{K_1\big(\frac{\vx^T\vbeta}{h}\big): \vbeta\in\bbB\subset\bbR^p, h\in(0,1]\big\}$. 
	We will first prove $\mathcal{F}_{10}$ is a VC class by similar techniques as in \citet{Sang2010} and \citet{Mason}. 
	It is sufficient to show the collection of all subgraphs
	$S_{0} =  \Big\{ \big\{(\vx,t): K_1\big(\frac{\vx^T\vbeta}{h}\big)<t\big\}: K_1\big(\frac{\vx^T\vbeta}{h}\big) \in\mathcal{F}_{10} \Big\}$
	forms a VC class of sets in $\mathcal{X}\times \bbR$.
	
	Since $K_1(\cdot)$ is a bounded, nondecreasing function, assume $\lim\limits_{x\ra-\infty} K_1(x) = m_1$ and $\lim\limits_{x\ra\infty}  K_1(x) = m_2$. Note that \vspace{-1em}
	\begin{align*}
	\Big\{(\vx,t): K_1\big(\frac{\vx^T\vbeta}{h}\big)<t\Big\} =\big \{(\vx,t): -\vx^T\vbeta+hK_1^{-1}(t)>0\big\},
	\end{align*} where 
	$K_1^{-1}(t) = -\infty$ if $t\leq m_1$, is $K_1^{-1}(t)$ for $m_1<t\leq m_2$ and is $\infty$ if $t>m_2$.
	Let $\psi_{\vbeta,h}(\vx,t) = -\vx^T\vbeta+hK_1^{-1}(t)$, $S_{1}=\{(\vx,t): \vx\in\mathcal{X},t\in(m_1,m_2]\}$ and $ S_{2}=\{(\vx,t): \vx\in\mathcal{X},t>m_2\}$. Then for any $\vbeta\in\bbB\subset\bbR^p$, $h\in\big(0,1\big]$,
	\begin{align*}
	&\Big\{(\vx,t): K_1\big(\frac{\vx^T\vbeta}{h}\big)<t\Big\} =\big \{(\vx,t): \psi_{\vbeta,h}(\vx,t) I((\vx,t)\in S_1)>0\big\}\cup S_2.
	\end{align*} 
	Note that $\psi_{\vbeta,h}(\vx,t) $ is in a finite dimensional space of functions when restricted to $S_1$. This implies the collection $\big\{(\vx,t):\psi_{\vbeta,h}(\vx,t) I((\vx,t)\in S_1)>0 \big\}$ is a VC subgraph class (Lemma 2.6.15, \citet{van1996weak}). $\{S_2\}$ is obviously VC. Hence, $S_0$ is also VC, and hence Donsker.
	As $(2A-1)Y$ is square integrable and does not depend on $(\vbeta, h)$, $\mathcal{F}_1$ is a Donsker class with a square integrable envelope (Theorem 2.10.6, \citet{van1996weak}). Similarly, $\mathcal{F}_2$ is also a Donsker class. Then by the Donsker presentation property, $\mathcal{F}$ is Donsker. 
\end{newproof}

\begin{lemma}
	Let $G_i(\vx_i,\vbeta,h_n)=(2A_i-1)K\big(\frac{\vx_i^T\vbeta}{h_n}\big)Y_i - \E \big\{(2A_i-1)\indicator(\vx_i^T\vbeta>0)Y_i\big\} $. 
	Under Assumptions \ref{A1}-\ref{A3} and \ref{K1},
	$\sup\limits_{\vbeta\in\bbB}\big|n^{-1}\sum_{i=1}^{n}G_i(\vx_i,\vbeta,h_n)\big|\rap 0$.
\end{lemma}

\begin{newproof}{ Lemma~\ref{Lemma5}}
	The Donsker property of $\mathcal{F}$ implies that as $n\ra\infty$, $$\sup\limits_{\vbeta\in\bbB}\Big|2n^{-1}\sum_{i=1}^{n}\Big[(2A_i-1)K\big(\frac{\vx_i^T\vbeta}{h_n}\big)Y_i - \E\big\{(2A_i-1)K\big(\frac{\vx_i^T\vbeta}{h_n}\big)Y_i\big\}\Big]\Big|\rap 0.$$
	It is sufficient to show $$\sup\limits_{\vbeta\in\bbB}\Big| \E \Big[2(2A_i-1)\big\{K\big(\frac{\vx_i^T\vbeta}{h_n}\big) - \indicator(\vx_i^T\vbeta>0)\big\}Y_i\Big]\Big|\ra 0.$$
	Note that 
	\bqa 
	&& \E \Big[2(2A_i-1)\big\{K\big(\frac{\vx_i^T\vbeta}{h_n}\big) - \indicator(\vx_i^T\vbeta>0)\big\}Y_i\Big]\\
	& =&\E \Big[\big\{K\big(\frac{\vx_i^T\vbeta}{h_n}\big)-\indicator(\vx_i^T\vbeta>0 ) \big\} \big\{Y_i^*(1)-Y_i^*(0)\big\}\Big]\\
	&=&\E \Big[\big\{K\big(\frac{\vx_i^T\vbeta}{h_n}\big)-\indicator(\vx_i^T\vbeta>0)\big\}\big\{\mu(1,\vx_i)- \mu(0,\vx_i)\big\}\Big].
	\eqa
	
	According to \ref{A1}, we know that $\mu(a,\vx)$ is bounded for almost all $\vx$, and $a=0,1$.
	\begin{align*}
	\sup\limits_{\vbeta\in\bbB}\Big| \E \Big[2(2A_i-1)\big\{K\big(\frac{\vx_i^T\vbeta}{h_n}\big) -\indicator(\vx_i^T \vbeta>0)\big\} Y_i\Big]\Big| 
	\leq\sup\limits_{\vbeta\in\bbB}c\E \Big|K\big(\frac{\vx_i^T\vbeta}{h_n}\big)- \indicator(\vx_i^T\vbeta>0)\Big|,
	\end{align*}
	for some positive constant $c$. For any positive constant $\tau$,
	\begin{align*}
	\sup\limits_{\vbeta\in\bbB}\Big| \E \Big[2(2A_i-1)\big\{K\big(\frac{\vx_i^T\vbeta}{h_n}\big) -\indicator(\vx_i^T \vbeta>0)\big\} Y_i\Big]\Big|\leq I_1+I_2,
	\end{align*}
	where
	\begin{align*}
	I_1 &= \sup\limits_{\vbeta\in\bbB}c\E\Big |\big\{K\big(\frac{\vx_i^T\vbeta}{h_n}\big)- \indicator(\vx_i^T\vbeta>0)\big\} \indicator(|\vx_i^T\vbeta|\geq \tau)\Big|,\\
	I_2 &= \sup\limits_{\vbeta\in\bbB}c\E \Big|\big\{K\big(\frac{\vx_i^T\vbeta}{h_n}\big)- \indicator(\vx_i^T\vbeta>0) \big\} \indicator(|\vx_i^T\vbeta|< \tau)\Big|.
	\end{align*}
	
	By the property of $K(\cdot)$, $\forall \tau>0$, the expectation can be made arbitrary small, uniformly in $\vbeta$, for all $n\geq n_0$, where $n_0$ is a positive integer. So $I_1\ra 0$. On the other hand,
	\begin{align*}
	I_2 \leq c\sup\limits_{\vbeta\in\bbB}P\big(|\vx_i^T\vbeta|< \tau\big) 
	= c\sup\limits_{\vbeta\in\bbB} P(-\tau-\vxw_i^T\vbetaw<x_1<\tau+\vxw_i^T\vbetaw) \leq c'\tau.
	\end{align*}
	As $\tau$ is an arbitrary positive constant, $I_2\ra 0$. This proves the lemma.
\end{newproof}

\begin{lemma}
	For any $\vtheta \in \bbR^{p-1}$, let
	$\vR_n(\vtheta) = \frac{2}{nh_n^2}\sum_{i=1}^n(2A_i-1)K'\big(\frac{z_i}{ h_n}+  \vtheta^T \vxw _i\big) \vxw _iY_i$.
	let $\eta >0$ be such that $S^{(1)}(z,\vxw )$, $S^{(2)}(z,\vxw )$, and $f^{(1)}(z|\vxw )$ exist and are uniformly bounded for almost 
	every $\vxw $ if $|z| \leq \eta$.  Define 
	$\Theta_n = \big\{\vtheta : \vtheta\in \bbR^{p-1}, h_n||\vtheta|| \leq \frac{\eta}{2\sqrt{p-1}M_x} \big\}$.
	Assume the conditions of Theorem~\ref{Th2} are satisfied, then
	(1) $\sup\limits_{\vtheta\in\Theta_n}||\vR_n(\vtheta)-\E \vR_n(\vtheta)||\rap 0 $.
	(2) There are finite numbers $\alpha_1$ and $\alpha_2$ such that for all $\vtheta\in\Theta_n$,
	$||\E \vR_n(\vtheta) - \vQ\vtheta|| \leq o(1)+\alpha_1h_n||\vtheta|| +\alpha_2h_n||\vtheta||^2$
	uniformly over  $\vtheta\in\Theta_n$.
\end{lemma}

\begin{newproof}{ Lemma~\ref{L3}} 
	(1) Let $k_{ni}(\vpsi)= (2A_i-1)K'\big(\frac{z_i+\vpsi^T\vxw_i}{h}\big)\vxw_i Y_i$. It suffices to show that $$\sup\limits_{\vpsi\in\Psi}\Big|\Big|(nh_n^2)^{-1}\sum_{i=1}^{n}\big\{k_{ni}(\vpsi)-\E k_{ni}(\vpsi)\big\}\Big|\Big|\rap 0,$$
	where $\Psi =\big\{\vpsi: \vpsi\in\bbR^{p-1}, ||\vpsi||\leq \frac{\eta}{2\sqrt{p-1}M_x}\big\}$. 
	
	The Donsker property of $\mathcal{H}$ implies that 
	$$\sup\limits_{\vpsi\in\Psi}\sup\limits_{h\in(0,1]} \Big|\Big|n^{-1}\sum_{i=1}^{n}\big\{k_{ni}(\vpsi)-\E k_{ni}(\vpsi)\big\}\Big|\Big|= O_p(n^{-1/2}).$$
	Then since $h_n \ra0$ and $n h_n^4\ra\infty$, we can derive that
	\begin{align*}
	\sup\limits_{\vpsi\in\Psi}\Big|\Big|(nh_n^2)^{-1}\sum_{i=1}^{n}\big\{k_{ni}(\vpsi)-\E k_{ni}(\vpsi)\big \}\Big|\Big|
	&\leq h_n^{-2}\sup\limits_{\vpsi\in\Psi}\sup\limits_{h\in(0,1]} \Big|\Big|n^{-1}\sum_{i=1}^{n} \big\{k_{ni}(\vpsi) -\E k_{ni}(\vpsi)\big\}\Big|\Big|\\
	&\leq O_p(n^{-1/2}h_n^{-2}) = o_p(1).
	\end{align*}
	
	(2)
	$\E \big\{\vR_n(\vtheta)\big\} = I_{n1}+I_{n2}$, where 
	$$I_{n1}=\frac{1}{h_n^2}\int_{|z|\leq\eta}K'\big(\frac{z}{ h_n}+ \vtheta^T \vxw \big) \vxw S(z,\vxw ) f(z|\vxw )dzdP(\vxw ),$$
	and $$I_{n2}=\frac{1}{h_n^2}\int_{|z|>\eta}K'\big(\frac{z}{ h_n}+ \vtheta^T \vxw \big) \vxw S(z,\vxw ) f(z|\vxw )dzdP(\vxw ).$$
	
	From \ref{A4} and \ref{A5}, we can say that for some $M>0$,
	$$||I_{n2}||\leq \frac{M}{h_n^2}\int_{|z|>\eta}K'\big(\frac{z}{ h_n}+ \vtheta^T \vxw \big)dzdP(\vxw ).$$
	Let $\zeta = z/h_n+  \vtheta^T \vxw $. Since $ h_n||\vtheta|| \leq \frac{\eta}{2\sqrt{p-1}M_x} $ and $||\vxw ||\leq \sqrt{p-1}M_x$ by \ref{A3}, then $|z|>\eta$ implies that $$|\zeta|>\frac{\eta}{2h_n} \mbox{,\quad and \quad }
	||I_{n2}||\leq \frac{M}{h_n}\int_{h_n|\zeta|>\eta/2}K'(\zeta)d\zeta.$$
	And from \ref{K3}, it converges to 0 as $n\ra \infty$. Therefore,
	$$\lim\limits_{n\ra \infty}\sup\limits_{\vtheta\in\Theta_n}||I_{n2}||=0.$$
	
	When $|z|\leq\eta$, then we have:
	\begin{align*} 
	S(z,\vxw ) f(z|\vxw ) = S^{(1)}(0,\vxw ) f(0|\vxw )z+\big\{S^{(1)}(0,\vxw ) f^{(1)}(\epsilon_2|\vxw ) + S^{(1)}(\epsilon_1,\vxw ) f^{(1)}(0|\vxw ) \big\}z^2,
	\end{align*}
	where $\epsilon_1$ and $\epsilon_2$  are between 0 and z. So $I_{n1}=J_{n1}+J_{n2}$, 
	where 
	\begin{align*} 
	J_{n1}&=\frac{1}{h_n^2}\int_{|z|\leq\eta}K'\big(\frac{z}{ h_n}+ \vtheta^T \vxw \big) \vxw zS^{(1)}(0,\vxw ) f(0|\vxw )dzdP(\vxw )\\
	&=\int_{|\zeta- \vtheta^T \vxw |\leq\eta/h_n}K'(\zeta)S^{(1)}(0,\vxw ) f(0|\vxw )\vxw (\zeta- \vxw^T  \vtheta)d\zeta dP(\vxw ),
	\end{align*}
	and
	\begin{align*}
	J_{n2}&=\frac{1}{h_n^2}\int_{|z|>\eta}K'\big(\frac{z}{ h_n}+ \vtheta^T \vxw\big) \vxw \big\{S^{(1)}(0,\vxw ) f^{(1)}(\epsilon_2|\vxw ) + S^{(1)}(\epsilon_1,\vxw ) f^{(1)}(0|\vxw ) \big\}z^2dzdP(\vxw )\\
	&=h_n\int_{|\zeta- \vtheta^T \vxw |>\eta/h_n}K'(\zeta) \vxw \big\{S^{(1)}(0,\vxw ) f^{(1)}(\epsilon_2|\vxw ) + S^{(1)}(\epsilon_1,\vxw ) f^{(1)}(0|\vxw ) \big\}(\zeta-\vxw^T  \vtheta )^2d\zeta dP(\vxw ).
	\end{align*}
	
	Since $\int \zeta K'(\zeta)d\zeta = 0$ by \ref{K2}, and $\big|h_n\vtheta^T \vxw \big|\leq \eta/2$,
	$$\Big|\int_{|\zeta- \vtheta^T \vxw |\leq\eta/h_n}\zeta K'(\zeta)d\zeta\Big|\leq \int_{|\zeta|\leq \eta/2h_n}\big|\zeta K'(\zeta)\big|d\zeta.$$
	
	By \ref{K2}, $\int_{|\zeta|\leq \eta/2h_n}\big|\zeta K'(\zeta)\big|d\zeta$ is bounded uniformly over $n$ and $\vtheta\in\Theta_n$, and converges to 0. So
	$$\lim\limits_{n\ra \infty}\sup\limits_{\vtheta\in\Theta_n}\Big|\Big|\int_{|\zeta- \vtheta^T \vxw |\leq\eta/h_n}\zeta K'(\zeta)S^{(1)}(0,\vxw ) f(0|\vxw )\vxw  d\zeta dP(\vxw )\Big|\Big|=0.$$
	
	In addition, 
	\begin{align*} 
	\Big|\vtheta^T \vxw -\vtheta^T \vxw \int_{|\zeta- \vtheta^T \vxw |\leq\eta/h_n}K'(\zeta) d\zeta \Big|&\leq \big|h_n\vtheta^T \vxw \big|h_n^{-1}\int_{|\zeta- \vtheta^T \vxw |\leq\eta/h_n}\big|K'(\zeta)\big|d\zeta \leq \frac{\eta}{2h_n}\int_{|\zeta|\geq\eta/(2h_n)}\big|K'(\zeta)\big|d\zeta.
	\end{align*}
	Similarly, we also have:
	$$\lim\limits_{n\ra \infty}	\Big|\Big|\sup\limits_{\vtheta\in\Theta_n}\int_{|\zeta- \vtheta^T \vxw |\leq\eta/h_n}\zeta K'(\zeta)S^{(1)}(0,\vxw ) f(0|\vxw )\vtheta^T \vxw \vxw^T d\zeta dP(\vxw ) -\vtheta^T \vQ	\Big|\Big|=0.$$
	
	Then for $J_{n2}$, there is some finite $M>0$, and $\alpha_1$, $\alpha_2$ such that:
	$$\big|\big|J_{n2}\big|\big|\leq Mh_n\int_{|\zeta- \vtheta^T \vxw |>\eta/h_n}\big|K'(\zeta)\big| (\zeta- \vtheta^T \vxw )^2d\zeta dP(\vxw ) \leq o(1)+\alpha_1h_n||\vtheta||+\alpha_2h_n||\vtheta||^2.$$
	
	In conclusion, 
	$\big|\big|\E \vR_n(\vtheta) - \vQ\vtheta\big|\big| \leq o(1)+\alpha_1h_n||\vtheta|| +\alpha_2h_n||\vtheta||^2.$
\end{newproof}

\begin{lemma}
	Define 
	$G^*_i(\vx_i,\vbeta,h_n)=(2A_i-1)r_iK\big(\frac{\vx_i^T\vbeta}{h_n}\big)Y_i - \E \big\{r_i(2A_i-1)\indicator(\vx_i^T\vbeta>0)Y_i\big\}$ .
	Under Assumptions \ref{A1}-\ref{A3} and \ref{K1},
	$\sup\limits_{\vbeta\in\bbB}\big|n^{-1}\sum_{i=1}^{n}G^*_i(\vx_i,\vbeta,h_n)\big| = \oprw$,
	where $\oprw$ denotes a random sequence that converges to zero in probability with respect to the joint distribution of $(r, w)$.
\end{lemma}

\begin{newproof}{ Lemma~\ref{Lemma10}}
	The Donsker property of $\mathcal{F}^*$ implies that as $n\ra\infty$, $$\sup\limits_{\vbeta\in\bbB}\Bigg|\frac{2}{n}\sum_{i=1}^{n}\Big[(2A_i-1)r_iK\big(\frac{\vx_i^T\vbeta}{h_n}\big)Y_i - \E_{w}\E_{r|w} \big\{(2A_i-1)r_iK\big(\frac{\vx_i^T\vbeta}{h_n}\big)Y_i\big\}\Big]\Bigg|= \oprw.$$
	It is sufficient to show 
	\begin{align*}
	&\sup\limits_{\vbeta\in\bbB}\Bigg| \E_{w}\E_{r|w} \Big[2(2A_i-1)r_i \big\{K\big(\frac{\vx_i^T\vbeta}{h_n}\big)- \indicator(\vx_i^T\vbeta>0)\big\}Y_i\Big]\Bigg|\\
	=&\sup\limits_{\vbeta\in\bbB}\Bigg| \E_{w} \Big[2(2A_i-1)\big\{K\big(\frac{\vx_i^T\vbeta}{h_n}\big)- \indicator(\vx_i^T\vbeta>0)\big\}Y_i\Big]\Bigg|\ra 0,
	\end{align*}
	which is verified in Lemma~\ref{Lemma5}. Hence the lemma is proved.
\end{newproof}

\begin{lemma} 
	Assume the conditions of Theorem~\ref{Th4} are satisfied, then
	$(nh_n)^{1/2}\big\{T_n^*(\vbetah_n; h_n)-T_n^*(\vbeta_0; h_n)\big\}=o_{p_r}(1)$,
	where $T_n^*(\vbeta,h_n)$ is defined as follows:
	\begin{align*}
	\vT_n^*(\vbeta; h_n) &= \frac{\partial \widetilde{M}^*_n(\vbeta, h_n)}{\partial \vbetaw} =  \frac{2}{n} \sum_{i=1}^n r_i(2A_i -1) K'\big(\frac{\vx_i^T\vbeta}{ h_n} \big) \frac{\vxw_i}{h_n} Y_i. 
	\end{align*}
\end{lemma}

\begin{newproof}{ Lemma~\ref{L9}}
	It follows from Lemma 3 of  \citet{ChengHuang} that
	it is sufficient to prove
	$$\sup\limits_{||\vbeta-\vbeta_0||\leq C(nh_n)^{-1/2}}\sqrt{nh_n}\big|\big|T_n^*(\vbeta; h_n)-T_n^*(\vbeta_0; h_n)\big|\big| = \oprw.$$
	According to Lemma 9.14 in \citet{kosorok2010introduction}, $\mathcal{H}^*$ is a bounded uniform entropy integral (BUEI) class, and the proof of  Lemma 9.13 implies that $\forall\ 0<\epsilon<1$, the $\epsilon$-covering number of $\mathcal{H}^*$ satisfies $N\big(\epsilon||F||,\mathcal{H}^*,L(P)\big) \leq \big(\frac{A}{\epsilon}\big)^v$, for some positive constants $A$ and $v$, and an envelop $F$. Consider the stochastic process $$f \mapsto n^{-1/2}\sum_{i=1}^n f(W_i,r_i),\quad f\in\mathcal{H}^*,\quad W_i = (A_i,\vx_i,Y_i).$$
	Given $Y,\vx, r$ all have sub-Gaussian distributions, $(f-\bbP f)$ is a separable sub-Gaussian process. Since $K''(\cdot)$ is bounded, we can derive by the Lipschitz property of $K'(\cdot)$ that
	$$||\vbeta-\vbeta_0||\leq C(nh_n)^{-1/2} \implies ||f-f_0||\leq C'(nh_n^3)^{-1/2}.$$ 
	By the property of the increments for the separable sub-Gaussian process (Corollary 2.2.8 in \citet{van1996weak}),
	\begin{align*}
	\E_{w}\E_{r|w}  &\Big[\sup\limits_{||f-f'||\leq C'(nh_n^3)^{-1/2}}\Big|\Big|n^{-1/2}\sum_{i=1}^n \big[ f(W_i,r_i) -\E_{w}\E_{r|w} f\big] - n^{-1/2}\sum_{i=1}^n \big\{f'(W_i,r_i)-\E_{w}\E_{r|w}  f'\big\}\Big|\Big| \Big]\\
	\leq&\ D \int_0^{C'(nh_n^3)^{-1/2}}\sqrt{\log(A/\epsilon)^v} d\epsilon
	\ \leq\  D' \big(nh_n^3\big)^{-1/2}\sqrt{\frac{1}{2}\log\big(nh_n^3\big)},
	\end{align*}
	for some positive constants $D$ and $D'$. Then by Markov inequlity, for any $\delta>0$,
	\begin{align*}
	&P_{r,w}\Big(\sup\limits_{||\vbeta-\vbeta_0||\leq C(nh_n)^{-1/2}}\sqrt{nh_n}\big|\big|T_n^*(\vbeta; h_n)-T_n^*(\vbeta_0; h_n)\big|\big|>\delta\Big)\\
	\leq&P_{r,w}\Big(\sup\limits_{||f-f_0||\leq C'(nh_n^3)^{-1/2}}h_n^{-1/2}\big|\big|n^{-1/2}\big(\sum_{i=1}^n f(W_i,r_i) - \sum_{i=1}^n f_0(W_i,r_i)\big)\big|\big|>\delta\Big)\\
	\leq&(\delta h_n^{1/2})^{-1}\E_{w}\E_{r|w}\Big[\sup\limits_{||f-f_0||\leq C'(nh_n^3)^{-1/2}}\big|\big|n^{-1/2}\big\{\sum_{i=1}^n f(w_i,r_i) - \sum_{i=1}^n f_0(W_i,r_i)\big\}\big|\big|\Big]\\
	\leq&(\delta h_n^{1/2})^{-1}\E_{w}\E_{r|w}\Big[\sup\limits_{||f-f_0||\leq C'(nh_n^3)^{-1/2}}n^{-1/2}\Big|\Big|\sum_{i=1}^n\big\{ f(W_i,r_i) -\E_w\E_{r|w} f\big\} - \sum_{i=1}^n \big\{f_0(W_i,r_i)-\E_w\E_{r|w} f_0\big\}\Big|\Big|\Big] \\
	& + (\delta h_n^{1/2})^{-1}\sup\limits_{||f-f_0||\leq C'(nh_n^3)^{-1/2}}\big|\big|n^{-1/2}\E_w\E_{r|w} (f-f_0)\big|\big|\\
	\leq& D' \delta^{-1}(nh_n^4)^{-1/2}\sqrt{\frac{1}{2}\log\big(nh_n^3\big)} +C' \delta^{-1}(nh_n^2)^{-1} \ra 0,
	\end{align*}
	given $\log (n) /nh_n^4=o_p(1)$, where $P_{r,w}(\cdot)$ denotes probability with respect to the joint distribution of $(r, w)$. The conclusion follows as $\delta> 0$ is arbitrary.	
\end{newproof}

\section{Proof of Theorems \ref{Th1}--\ref{Th4}}\label{Proofs}
\begin{newproof}{ Theorem~\ref{Th1}}
	We observe that $\vbetah_n$ maximizes $\widetilde{M}_n(\vbeta, h_n)$ over $\vbeta\in \bbB$.
	Lemma~\ref{Lemma5} implies that
	$	\sup\limits_{\vbeta  \in \bbB} \big|\widetilde{M}_n(\vbeta, h_n) -  M(\vbeta)\big| \ra 0$ in probability as $n\ra \infty$, where $M(\vbeta)=\E \{(2A_i-1)\indicator(\vx_i^T\vbeta>0)Y_i\}.$ Condition \ref{A3}  implies that for every $\epsilon>0$, 
	$\sup_{||\vbeta-\vbeta_0|| > \epsilon}M(\vbeta)<M(\vbeta_0)$. Hence, $\vbetah_n$ is consistent by Theorem 5.7 in \citet{van2000asymptotic}.  
\end{newproof} 


The asymptotic distribution of $\vbetaw_n$ depends critically on the properties of the gradient and the Hessian matrix of the objective function. $ \widetilde{M}_n(\vbeta, h_n)$. Define
\begin{align}
\vT_n(\vbeta; h_n) &= \frac{\partial \widetilde{M}_n(\vbeta, h_n) }{\partial \vbetaw} =  \frac{2}{n}\sum_{i=1}^n (2A_i -1) K'\big(\frac{\vx_i^T\vbeta}{ h_n} \big) \frac{\vxw_i}{h_n}Y_i,\label{Tn}\\
\vQ_n(\vbeta; h_n) &= \frac{\partial ^2  \widetilde{M}_n(\vbeta, h_n)}{\partial \vbetaw \partial \vbetaw^T}  =  \frac{2}{n}\sum_{i=1}^n (2A_i -1) K''\big(\frac{\vx_i^T\vbeta}{ h_n} \big) \frac{\vxw_i\vxw_i^T}{h_n^2}Y_i.	 \label{Qn}
\end{align}

\setcounter{lemma}{1}
\renewcommand{\thelemma}{S\arabic{lemma}}

Lemmas~\ref{L1} and \ref{L45} below establish useful properties of $\vT_n(\vbeta; h_n)$ and $\vQ_n(\vbeta; h_n)$, respectively. The proofs of these two lemmas are given in Section~\ref{Smoothproof}.  
\begin{lemma}
	\label{L1}
	Assume the conditions of Theorem~\ref{Th2} are satisfied, then
	\bqa
	\lim\limits_{n \ra \infty} \E\{ (nh_n)^{1/2}\vT_n(\vbeta_0; h_n)\}=0,\mbox{\quad and \quad}\lim\limits_{n \ra \infty} \Var\{(nh_n)^{1/2}\vT_n(\vbeta_0; h_n) \}=\vD.
	\eqa
\end{lemma}

\begin{lemma}
	\label{L45}  
	Let $\vbeta_{n}^r$ be any value between $\vbetah_n$ and $\vbeta_0$. 
	Assume the conditions of Theorem~\ref{Th2} are satisfied, then
	$\vQ_n(\vbeta_n^r; h_n) \rap \vQ$, where $\vQ$ is defined in (\ref{vQ}).		
\end{lemma}

Let $V''(\cdot)$ be the Hessian matrix of $V(\cdot)$ with respect to $\vbetaw$, i.e., $ \frac{\partial^2 V(\vbeta)}{\partial\vbetaw\partial\vbetaw^T}$. Lemmas~\ref{Hessian}  below describes the continuity of  $V''(\cdot)$. The proof is  given in Section~\ref{Smoothproof}.
\begin{lemma}
	\label{Hessian}  
	Let $\vbeta_{n}^r$ be any value between $\vbetah_n$ and $\vbeta_0$. 
	Assume the conditions of Theorem~\ref{Th2} are satisfied, then $V''(\vbeta_{n}^r)  \rap \vI_V$, where $\vI_V = -\E \big\{S^{(1)}(0,\vxw)f(0|\vxw)\vxw\vxw^T(\vxw^T\vbetaw_0)\big\}$.
\end{lemma}

\begin{newproof}{ Theorem~\ref{Th2}}
	By Taylor expansion, we have
	$$\vT_n(\vbetah_n; h_n) = \vT_n(\vbeta_0; h_n) +  \vQ_n(\vbeta_n^r; h_n)(\vbetaw_n  - \vbetaw_0),$$
	where $\vbeta_{n}^r$ is between $\vbetah_n$ and $\vbeta_0$. 
	The definition of $\vbetah_n$ implies that $ \vT_n(\vbetah_n; h_n) = \widetilde{\vnull}$, where 
	$\widetilde{\vnull}$ denotes a $(p-1)$-dimensional vector of zeroes.
	Lemma~\ref{L45} indicates that 
	$$\vbetaw_n -\vbetaw_0=\big(-\vQ+o_p(1)\big)^{-1}\vT_n( \vbeta_0; h_n).$$
	To prove the theorem, it suffices to verify $(nh_n)^{1/2}\vT_n(\vbeta_0; h_n) \rad N(0, \vD)$.
	It is known from Lemma B1 that $\E (nh_n)^{1/2}\big\{\vT_n(\vbeta_0; h_n) \big\} \ra 0$. It is sufficient to prove that 
	$(nh_n)^{1/2}\vgamma^T\big\{\vT_n(\vbeta_0; h_n)-\E \vT_n(\vbeta_0; h_n)\big\}$ is asymptotically 
	$N(\widetilde{\vnull},\vgamma^T\vD\vgamma)$ for any constant vector $\vgamma\in\bbR^{p-1}$ such that $||\vgamma||=1$. Let
	$$q_i = 2(2A_i-1)(nh_n)^{1/2}K'\big(\frac{\vx_i^T\vbeta_0}{ h_n} \big) \frac{\vgamma^T\vxw _i}{h_n}Y_i.$$
	It follows the proof of Lemma~\ref{L1} that $\lim\limits_{n \ra \infty} \E q_i= 0$, and $\lim\limits_{n \ra \infty} \E q_i^2/n=  \vgamma^T\vD\vgamma$.
	
	To apply Lyapunov central limit theorem, we will verify that 
	\begin{align}
	\lim\limits_{n \ra \infty} (s_n^4)^{-1} \sum_{i=1}^n \E \big\{(q_i-\E q_i)^4\big\}=0,
	\label{dolphin}
	\end{align}
	where
	$\lim\limits_{n \ra \infty}(n^{-2}s_n^2)= \lim\limits_{n \ra \infty}\sum_{i=1}^n \Var(n^{-1}q_i)= \vgamma^T\vD\vgamma$.
	We observe that the left-side of (\ref{dolphin}) is bounded from above (up to a positive constant) by
	$\lim\limits_{n \ra \infty} n^{-3} \E (q_i^4) +\lim\limits_{n \ra \infty}n^{-3}(\E q_i)^4 = I_1+I_2$.
	As $(\E q_i)^4 \ra  0$,  we have $I_2=o(1)$. To evaluate $I_1$, note that 
	$$
	n^{-3}\E (q_i^4) = 16(nh_n^2)^{-1}\E \big\{K'\big(\frac{x_i^T\vbeta}{ h_n}\big)^4( \vgamma^T\vxw _i)^4 Y_i^4\big\}.
	$$
	Since $Y$ has sub-Gaussian tail, then for any integer $k \geq 1$, $\E|Y|^k$ is finite. So with the boundedness of $K(\cdot)$ and $\vxw$, $\E \big\{K'\big(\frac{x_i^T\vbeta}{ h_n}\big)^4( \vgamma^T\vxw _i)^4 Y_i^4\big\}$ is finite. Then $n^{-1}h_n^{-2}= o(1)$ implies $I_1=o(1)$.  Therefore, the Lyapunov condition is satisfied.  This proves (1). 
	
	To prove (2), we observe that 
	\begin{align*}
	\sqrt{n}(V_n(\vbetah_n)-V(\vbeta_0))=&\sqrt{n}\big\{V_n(\vbeta_0)-V(\vbeta_0)\big\} +\sqrt{n}\big\{V_n(\vbetah_n)-V_n(\vbeta_0)-V(\vbetah_n)+V(\vbeta_0)\big\} \\&+\sqrt{n}\big\{V(\vbetah_n)-V(\vbeta_0)\big\}\\
	=&I_1+I_2+I_3,
	\end{align*} 
	where the definition of $I_i$ ($i=1, 2, 3$) is clear from the context. 
	By the central limit theorem, we have $I_1\rad N(0,U)$. The Donsker property of the function class $\mathcal{G}$ ensures that
	$I_2=o_p(1)$.
	Note that with the consistency result in Theorem \ref{Th1}, we have $P(\vbetah_{n1}=\vbeta_{01})\ra 1$ as $n\ra \infty$. By Taylor expansion, we have\vspace{-0.1em}
	$$	I_3=\sqrt{n}(\vbetaw_n-\vbetaw_0)^TV'(\vbeta_0) +\sqrt{n}(\vbetaw_n-\vbetaw_0)^TV''(\vbeta^r)(\vbetaw_n-\vbeta_0)/2+o_p(1),$$
	where $V'(\cdot)$ and $V''(\cdot)$ denote the gradient vector and Hessian matrix of $V(\cdot)$ with respect to $\vbetaw$, respectively; 
	$\vbeta^r$ is between $\vbetaw_0$ and $\vbetaw_n$. As $\vbeta_0$ is the maximizer of $V(\cdot)$,
	we have $V'(\vbeta_0)=0$. Let $\lambda_{max}(\cdot)$ be the eigenvalue with the greatest absolute value. The second term is 
	upper bounded 
	by $|\lambda_{max}(V''(\vbeta^r))|\sqrt{n}||\vbetaw_n-\vbetaw_0||^2/2$,
	which is of order $O_p(n^{-1/2}h^{-1})=o_p(1)$
	by Lemma~\ref{Hessian}, Assumption \ref{A5} and the first part of the theorem on the convergence rate.	
	This proves (2). 
\end{newproof}

In the rest of this appendix, we will prove the theory for bootstrap based inference. As described in Section~\ref{Theory2}, given a sequence of random variables $R_n$, $n=1, \ldots, n$, we write $R_n=\opr$ if for any $\epsilon>0, \delta>0$,  we have $P_{w}\big(P_{r|w}(|R_n|>\epsilon)>\delta\big)\ra 0$ as $n\ra \infty$. 
Furthermore, $\oprw$ denotes a random sequence that converges to zero in probability with respect to the joint distribution of $(r, w)$; and $\opw$ denotes a random sequence that converges to zero in probability with respect to the distribution of $r$ only.
By Lemma 3 of  \citet{ChengHuang}, if $R_n=o_{p_{rw}}(1)$, then  $R_n=o_{p_{r}}(1)$.
In particular, if  $R_n$ depends only on the data $w$ but not on the random weights $r$ and if $R_n=o_{p_{w}}(1)$, then it is easy to see $R_n=o_{p_{rw}}(1)$, and hence it is $o_{p_{r}}(1)$.
In this part of proof, we will include subscripts in the probability and expectation to clarify which probability distribution is used in the calculation.\vspace{1em}

\begin{newproof}{ Theorem~\ref{Th3}}
	By definition, $\vbetah^*_n$ maximizes $\widetilde{M}^*_n(\vbeta, h_n)$ over $\vbeta\in \bbB$.
	First, by combining Lemma~\ref{Lemma5} and Lemma~\ref{Lemma10} and recognizing that $\E_w \{(2A_i-1)\indicator(\vx_i^T\vbeta>0)Y_i\}
	=\E_w\E_{r|w} \big\{r_i(2A_i-1)\indicator(\vx_i^T\vbeta>0)Y_i\big\}$, we have
	$\sup\limits_{\vbeta  \in \bbB}|\widetilde{M}^*_n(\vbeta, h_n) -\widetilde{M}_n(\vbeta, h_n) | = \oprw$.  By Lemma~3 of \citet{ChengHuang},
	$\sup\limits_{\vbeta  \in \bbB}  \big|\widetilde{M}^*_n(\vbeta, h_n) -\widetilde{M}_n(\vbeta, h_n) \big|  =\opr \pxp$.
	By Theorem 5.7 in \citet{van2000asymptotic}, to prove the theorem, it is sufficient to show that for any $\epsilon > 0$,   \vspace{-1em}
	\bqan\label{bear}
	\lim\limits_{n\ra\infty}P_{w}\Big(\sup\limits_{||\vbeta-\vbetah_n||> \epsilon} 
	\big\{ \widetilde{M}_n(\vbeta, h_n) - \widetilde{M}_n(\vbetah_n, h_n) \big\}<0\Big)= 1.
	\eqan 
	By Lemma A2,
	$\widetilde{M}_n(\vbeta, h_n) - \widetilde{M}_n(\vbetah_n, h_n)=M(\vbeta)-M(\vbetah_n)+o_{p_{w}}(1)$.
	Furthermore, the consistency of $\vbetah_n$ implies that for all sufficiently large $n$, any $\vbeta$ that satisfies $||\vbeta- \vbetah_n||> \epsilon$ would also satisfy $||\vbeta-\vbeta_0||\geq \epsilon/2$. 
	Condition \ref{A3}  implies that $\sup_{||\vbeta-\vbeta_0|| > \epsilon/2}M(\vbeta)<M(\vbeta_0)$. Hence, (\ref{bear}) holds. This proves (1). 
	
	To prove (2), we observe that 
	$\sqrt{n}(V_n^*(\vbeta)-V_n(\vbeta))=n^{-1/2}\sum_{i=1}^n(r_i-1)\big\{A_i\indicator(\vx_i^T\vbeta>0)+ (1-A_i)\indicator(\vx_i^T\vbeta\leq 0)\big\}Y_i$,
	which has mean zero. The Donsker property of the function class $\mathcal{G}^*$ and the fact $\vbetah_n=\vbeta_0+\oprw$ implies that
	\begin{align}
	\sqrt{n}\big[\{V_n^*(\vbetah_n)-V_n(\vbetah_n)\}-\{V_n^*(\vbeta_0)-V_n(\vbeta_0)\}\big]= \oprw,\label{value3}
	\end{align} 
	by	Lemma 19.24 of \citet{van2000asymptotic}. By assumption~\ref{A6} and the classical central limit theorem, $\sqrt{n}\{V_n^*(\vbeta_0)-V_n(\vbeta_0)\}= N(0,U)+\oprw$.
	Hence, $\sqrt{n}\big\{V_n^*(\vbetah_n)-V_n(\vbetah_n)\big\}= N(0,U)+\oprw$. Lemma~3 in \citet{ChengHuang} implies (2) holds.
\end{newproof}

To prove Theorem ~\ref{Th4}, we define the following gradient function and Hessian matrix corresponding to the randomly weighted objective function
\begin{align}
\vT_n^*(\vbeta; h_n) &= \frac{\partial \widetilde{M}^*_n(\vbeta, h_n)}{\partial \vbetaw} =  \frac{2}{n} \sum_{i=1}^n r_i(2A_i -1) K'\big(\frac{\vx_i^T\vbeta}{ h_n} \big) \frac{\vxw_i}{h_n} Y_i,\label{Tn*}\\
\vQ_n^*(\vbeta; h_n) &= \frac{\partial ^2  \widetilde{M}^*_n(\vbeta, h_n)}{\partial \vbetaw \partial \vbetaw^T}  =  \frac{2}{n}\sum_{i=1}^n r_i(2A_i -1) K''\big(\frac{\vx_i^T\vbeta}{ h_n} \big) \frac{\vxw_i\vxw_i^T}{h_n^2}Y_i.	\label{Qn*}
\end{align}
Lemma \ref{L78} below characterizes the asymptotic property of the Hessian matrix. Its proof is given in the supplementary material. 
\begin{lemma}
	\label{L78}
	 Let $\vbeta_{n}^{*r}$ be a variable between $\vbetah_n^*$ and $\vbetah_n$.
	Assume the conditions of Theorem~\ref{Th4} are satisfied, then
	$\vQ_n^*(\vbeta_{n}^{*r}; h_n) = \vQ + \opr\pxp$.
\end{lemma}

\begin{newproof}{ Theorem~\ref{Th4}}
	By Taylor expansion, 
	$\vT^*_n(\vbetah_n^*; h_n) = \vT^*_n(\vbetah_n; h_n) +  \vQ^*_n(\vbeta_n^{*r}; h_n)(\vbetaw_{n}^{*}-\vbetaw_n)$,
	where $\vbeta_{n}^{*r}$ is between $\vbetah_n^*$ and $\vbetah_n$. By the definition of $\vbetah_n^*$, we have 
	$ \vT_n^*(\vbetah_n^*; h_n) = \widetilde{\vnull}$. By Lemma B4, we have 
	$$\vbetaw_{n}^{*}-\vbetaw_n = -\big(\vQ + \opr\big)^{-1}\vT^*_n(\vbetah_n; h_n).$$
	It remains to show $(nh_n)^{1/2}\vT_n^*(\vbetah_n; h_n) = N(0, \vD)+\opr$. By Lemma A5, we only need to show $(nh_n)^{1/2}\vT_n^*(\vbeta_0; h_n) = N(0, \vD)+\opr$.
	Observe that
	\begin{align*} 
	\E_{r|w} \big\{(nh_n)^{1/2}\vT_n^*(\vbeta_0; h_n) \big\}& = (nh_n)^{1/2}\vT_n(\vbeta_0; h_n), \\
	\Var_{r|w} \big\{(nh_n)^{1/2}\vT_n^*(\vbeta_0; h_n)\big\} &=\frac{4 }{n}\sum_{i=1}^n \big\{K'\big(\frac{\vx_i^T\vbeta_0}{ h_n} \big) \big\}^2 \frac{\vxw_i\vxw_i^T}{h_n}Y_i^2.
	\end{align*}
	Lemma~\ref{L1} implies that 
	$$\lim\limits_{n\ra \infty}\E_w\E_{r|w}  \big\{(nh_n)^{1/2}\vT_n^*(\vbeta_0; h_n) \big\} = 0\mbox{, \quad and \quad }\lim\limits_{n\ra \infty}\E_w\Big[\Var_{r|w} \big\{(nh_n)^{1/2}\vT_n^*(\vbeta_0; h_n)\big\}  \Big] = \vD.$$ 
	It suffices to prove that for any constant vector $\vgamma\in\bbR^{p-1}$ such that $||\vgamma||=1$,
	\bqa
	(nh_n)^{1/2}\vgamma^T\big\{\vT^*_n(\vbeta_0; h_n)-\E \vT^*_n(\vbeta_0; h_n)\big\}=
	N(\widetilde{\vnull},\vgamma^T\vD\vgamma) +\opr.
	\eqa
	Define
	$q_i^* =2r_i(2A_i-1)(nh_n)^{1/2}K'\big(\frac{\vx_i^T\vbeta_0}{ h_n} \big) \frac{\vgamma^T\vxw _i}{h_n}Y_i,$
	where $\E_{r|w} q^*_i=  q_i$, and $\E_{r|w} (q_i^{*2})=  2q_i^2$, for $q_i$ defined in the proof of Theorem~\ref{Th2}. To check the Lyapunov condition, it suffices to prove that 
	$$	(s_n^{*4})^{-1} \sum_{i=1}^n \E_{r|w} \big\{(q_i^*-\E_{r|w} q^*_i)^4\big\}\xrightarrow{a.s.} 0,$$
	where $s_n^{*2} = \sum_{i=1}^n\Var_{r|w}(q_i^*) $. Similarly as  Theorem 2, the Lyapunov condition holds if $$(s_n^{*4})^{-1}  \sum_{i=1}^n \E_{r|w} (q_i^{*4} )\xrightarrow{a.s.} 0\mbox{, \quad and \quad}(s_n^{*4})^{-1}  \sum_{i=1}^n  (\E_{r|w}q_i^*)^4 \xrightarrow{a.s.} 0. $$
	
	Since $r$ and $Y$ both have sub-Gaussian tails, we know that for any integer $k \geq 1$, $\E|r|^k$ and $\E|Y|^k$ are finite. Then it is easy to compute that $s_n^{*2} = 4n^2h_n^{-1}I_1$, $\sum_{i=1}^n \E_{r|w} (q_i^{*4} )= 16n^3h_n^{-2}\E(r^4)I_2$, and $\sum_{i=1}^n  (\E_{r|w}q_i^*)^4 =16n^3h_n^{-2}I_2 $, where
	\bqa
	I_1 = n^{-1} \sum_{i=1}^n\big\{K'\big(\frac{\vx_i^T\vbeta_0}{ h_n} \big)^2 (\vgamma^T\vxw _i)^2 Y_i^2\big\},\qquad I_2=n^{-1} \sum_{i=1}^n\big\{K'\big(\frac{\vx_i^T\vbeta_0}{ h_n} \big)^4 (\vgamma^T\vxw _i)^4Y_i^4\big\}.
	\eqa
	According to \ref{K1} and \ref{A1}-\ref{A2}, we know that $I_1$ and $I_2$ are both absolutely integrable. Then  the strong law of large numbers implies that $I_1\xrightarrow{a.s.}\E_w I_1$ and $I_2\xrightarrow{a.s.}\E_w I_2$. With the continuous mapping theorem, it is easy to conclude that $I_1^{-2}I_2\xrightarrow{a.s.}(\E_w I_1)^{-2}\E_w I_2$. We therefore have 
	\begin{align*}
	(s_n^{*4})^{-1} \sum_{i=1}^n \E_{r|w} (q_i^{*4} ) &=n^{-1}\E(r_i^4)I_1^{-2}I_2\xrightarrow{a.s.} 0,\quad
	(s_n^{*4})^{-1} \sum_{i=1}^n (\E_{r|w}q_i^*)^4  = n^{-1}I_1^{-2}I_2\xrightarrow{a.s.} 0. 
	\end{align*}
	This verifies the Lyapunov condition and finishes the proof. 
\end{newproof}

\section{Proof of Auxiliary Results in Section~\ref{Proofs}}  \label{Smoothproof}
\begin{newproof}{ Lemma~\ref{L1}}
	(1)  Let $ \zeta = z/h_n$, then by \ref{A1}, we have
	\begin{align*} 
	\E \big\{h_n^{-b}\vT_n(\vbeta_0; h_n)\big\}&= h_n^{-b}\E \Big\{(2A -1) K'\big(\frac{x^T\vbeta_0}{ h_n} \big) \frac{\vxw }{h_n}Y\Big\} \\
	&=h_n^{-b} \E \Big\{ K'\big(\frac{x^T\vbeta_0}{ h_n} \big) \frac{\vxw }{h_n}(Y^*_1-Y^*_0)\Big\}\\
	&=h_n^{-b} \E \Big\{ K'\big(\frac{z}{ h_n} \big) \frac{\vxw }{h_n}S(z,\vxw )\Big\}\\
	&=h_n^{-b} \int K'(\zeta )\vxw S(h_n\zeta,\vxw )f(h_n\zeta|\vxw )d\zeta dP(\vxw).
	\end{align*}
	
	Under \ref{A3}, $S(0,\vxw ) =0$ for almost every $\vxw $, so the Taylor series expansions for $S(h_n\zeta,\vxw )$ and $f(h_n\zeta|\vxw )$ can be written as
	\begin{align*}
	S(h_n\zeta,\vxw ) &= \sum_{i=1}^{b-1}S^{(i)}(0,\vxw )\frac{(h_n\zeta)^i}{i!} +S^{(b)}(\xi_b,\vxw )\frac{(h_n\zeta)^b}{b!},\\
	f(h_n\zeta|\vxw ) &=\sum_{j=0}^{b-i-1}f^{(j)}(0|\vxw )\frac{(h_n\zeta)^{j}}{(j)!} +  f^{(b-i)}(\xi_i|\vxw )\frac{(h_n\zeta)^{b-i}}{(b-i)!},
	\end{align*}
	for $i = 1,..., b-1$, where $\xi_1,...,\xi_b$ are scalars with values between $0$ and $h_n\zeta$. Combining these two expansions yields
	
	\begin{align}
	\begin{split}
	\label{Sp} 
	S(h_n\zeta,\vxw )f(h_n\zeta|\vxw ) =& S^{(b)}(\xi_b,\vxw )\frac{(h_n\zeta)^b}{b!}f(h_n\zeta|\vxw )
	+ \sum_{i=1}^{b-1}S^{(i)}(0,\vxw )f^{(b-i)}(\xi_i|\vxw )\frac{(h_n\zeta)^b}{i!(b-i)!}\\
	&+ \sum_{i=1}^{b-1}\sum_{j=0}^{b-i-1}S^{(i)}(0,\vxw )f^{(j)}(0|\vxw )\frac{(h_n\zeta)^{i+j}}{i!j!}.
	\end{split}
	\end{align}
	
	So we have
	\begin{align*} 
	\E \big\{h_n^{-b}\vT_n(\vbeta_0; h_n) \big\}=& \int \zeta^bK'(\zeta )\vxw \Big\{S^{(b)}(\xi_b,\vxw )\frac{f(h_n\zeta|\vxw )}{b!}+ \sum_{i=1}^{b-1}S^{(i)}(0,\vxw )\frac{f^{(b-i)}(\xi_i|\vxw )}{i!(b-i)!}\Big\} d\zeta dP(\vxw )\\
	&+   \sum_{i=1}^{b-1}\sum_{j=0}^{b-i-1}\int h_n^{i+j-b}\zeta^{i+j}K'(\zeta )\vxw S^{(i)}(0,\vxw )\frac{f^{(j)}(0|\vxw )}{i!j!}d\zeta dP(\vxw )\\
	&= I_1+I_2+I_3,
	\end{align*}
	where for some $\eta>0$,
	\begin{align*} 
	I_1&= \int \zeta^bK'(\zeta )\vxw \Big\{ S^{(b)}(\xi_b,\vxw )\frac{f(h_n\zeta|\vxw )}{b!}+ \sum_{i=1}^{b-1}S^{(i)}(0,\vxw )\frac{f^{(b-i)}(\xi_i|\vxw )}{i!(b-i)!}\Big\}d\zeta dP(\vxw ),\\
	I_2&= \sum_{i=1}^{b-1}\sum_{j=0}^{b-i-1}\int_{|h_n\zeta|\leq\eta}h_n^{i+j-b}\zeta^{i+j}K'(\zeta )\vxw S^{(i)}(0,\vxw )\frac{f^{(j)}(0|\vxw )}{i!j!}d\zeta dP(\vxw ),\\
	I_3&=\sum_{i=1}^{b-1}\sum_{j=0}^{b-i-1} \int_{|h_n\zeta|>\eta} h_n^{i+j-b} \zeta^{i+j}K'(\zeta )\vxw S^{(i)}(0,\vxw )\frac{f^{(j)}(0|\vxw )}{i!j!}d\zeta dP(\vxw ).
	\end{align*} 
	
	Then from \ref{K2}, \ref{A4}-\ref{A5}, and  Lebesgue's dominated convergence theorem, we have $I_1\ra \vH$,
	where 
	$\vH =a_{H} \sum_{i=1}^{b}\frac{1}{{i!(b-i)!}}\E\big\{\vxw S^{(i)}(0,\vxw)f^{(b-i)}(0|\vxw )\big\}$ with $a_{H} = \int \nu^bK'(\nu ) d\nu$. Similarly, let $\eta\ra 0$, we have $I_2\ra 0$ and $I_3\ra 0$. 
	Therefore, $\lim\limits_{n \ra \infty} \E \big\{(nh_n)^{1/2}\vT_n(\vbeta_0; h_n) \big\}=0$, since $nh_n^{2b+1}=o(1)$.
	
	
	(2) Let $t_i = 2(2A_i -1) K'\big(\frac{x_i^T\vbeta_0}{ h_n} \big) \frac{\vxw _i}{h_n}Y_i$, then we have:	
	\bqa
	\Var\big\{(nh_n)^{1/2}\vT_n \big\} &=& h_n \big(\E t_it_i^T-\E \vT_n\E \vT_n^T\big) \\
	&=&4\E   \Big\{K'\big(\frac{x_i^T\vbeta_0}{ h_n} \big) \Big\}^2\frac{\vxw _i\vxw _i^T}{h_n}Y_i^2-h_n\E \vT_n\E \vT_n^T\\
	&= &2\E \Big\{K'\big(\frac{z}{ h_n} \big)^2 \frac{\vxw \vxw^T}{h_n}\E \big(Y_1^{*2}+Y_0^{*2}|z,\vxw \big)\Big\}-h_n\E \vT_n\E \vT_n^T\\
	&= &2\int  K'\big(\frac{z}{ h_n} \big)^2 \frac{\vxw \vxw^T}{h_n}\E \big(Y_1^{*2}+Y_0^{*2}|z,\vxw \big)f(z|\vxw )dz dP(\vxw ) -h_n\E \vT_n\E \vT_n^T\\
	&= &2\int  K'(\zeta )^2 \vxw \vxw^T\E \big(Y_1^{*2}+Y_0^{*2}|h_n\zeta,\vxw \big)f(h_n\zeta|\vxw )d\zeta dP(\vxw ) -h_n\E \vT_n\E \vT_n^T.
	\eqa
	
	Since $h_n \ra 0$,  by \ref{A1}-\ref{A2} and the dominated convergence theorem, we have
	\begin{align*} 
	&\quad \lim\limits_{n \ra \infty} \big\{\Var(nh_n)^{1/2}\vT_n\big\} \\&= \lim\limits_{n \ra \infty}\Big\{2\int  K'(\zeta )^2 \vxw \vxw^T\E \big(Y_1^{*2}+Y_0^{*2}|h_n\zeta,\vxw\big) f(h_n\zeta|\vxw )d\zeta dP(\vxw )-h_n\E \vT_n\E \vT_n^T\Big\}\\
	&= 2\int  K'(\zeta )^2 d\zeta \int\vxw \vxw^T\E \big(Y_1^{*2}+Y_0^{*2}|0,\vxw\big) f(0|\vxw ) dP(\vxw )-0*\vH\vH^T\\
	&=a_1\int\vxw \vxw^Tf(0|\vxw )\E \big(Y_1^{*2}+Y_0^{*2}|0,\vxw\big) dP(\vxw ) =\vD.
	\end{align*}
	This finishes the proof.
\end{newproof}

\begin{newproof}{ Lemma~\ref{L45}}
	It suffices to prove the following two results:\\
	(a) let  $\vtheta_n = (\vbetah_n-\vbeta_0)/h_n$, then $\vtheta_n =o_p(1)$;\\
	(b) let $\{\vbeta_n\} = \{(\beta_{n1}, \vbetaw_n^T)^T\}$ be any sequence in $\bbB$ such that $(\vbeta_n-\vbeta_0)/h_n\ra 0$ as $n\ra\infty$, then $\vQ_n(\vbeta_n; h_n) \rap \vQ.$
	
	To prove (a), we first note that
	$h_n\vtheta_n\rap 0$ by Theorem~\ref{Th1}.  Lemma~\ref{L3} then implies $\vR_n(\vtheta_n)\rap 0$, 
	and there exist some constants $\alpha_1$ and $\alpha_2$ such that:
	$||\vQ \vtheta_n||\leq o_p(1)+\alpha_1h_n||\vtheta_n|| +\alpha_2h_n||\vtheta_n||^2$.
	Since $\vQ$ is negative definite, we have $\inf\limits_{\vtheta}\frac{||\vQ \vtheta||}{||\vtheta|| }= |\omega_{min}|>0$,  
	where $\omega_{min}$ is the eigenvalue of $\vQ$ with the smallest absolute value. It indicates that
	$$0< |\omega_{min}|<\frac{ ||\vQ \vtheta_n||}{||\vtheta_n||}\leq o_p(||\vtheta_n||^{-1})+\alpha_1h_n +\alpha_2h_n||\vtheta_n||.$$
	Since $h_n\ra 0$ and $h_n||\vtheta_n||\rap 0$, if $||\vtheta_n||= o_p(1)$ does not hold, then the right hand side of the above inequality would 
	degenerate to $o_p(1)$, which contradicts with the fact that it should be larger than $|\omega_{min}|>0$. Consequently, we 
	have $||\vtheta_n||= o_p(1)$.

	To prove (b),  let $q_{ni}(\vbeta)= (2A_i-1)K''\big(\frac{\vx_i^T\vbeta}{h}\big)\vxw_i\vxw_i^T Y_i$. It suffices to show that $$\sup\limits_{||\vbeta-\vbeta_0||\leq \epsilon h_n}\Big|\Big|n^{-1}\sum_{i=1}^{n} \big\{h_n^{-2}q_{ni}(\vbeta) - Q\big\}\Big|\Big|=o_p(1),$$
	for arbitrary positive $\epsilon$.
	
	First, let $\vbetaw_n=\vbetaw_0+h_n\tilde{\vtheta}_n$ with $\widetilde{\vtheta}_n\ra0$ as $n\ra\infty$. Now we have
	\begin{align*} 
	\E \big\{ h_n^{-2}q_{ni}(\vbeta_n) \big\} &= \E \Big\{2(2A -1) K''\big(\frac{x^T\vbeta_n}{ h_n} \big) \frac{\vxw \vxw ^T}{h_n^2}Y\Big\} \\
	&=\E \Big\{ K''\big(\frac{x^T\vbeta_0}{ h_n} + \widetilde{\vtheta}_n^T \vxw \big) \frac{\vxw \vxw '}{h_n^2}S(z,\vxw )\Big\}\\
	&=\int K''\big(\frac{z}{ h_n}+ \widetilde{\vtheta}_n^T \vxw \big) \frac{\vxw \vxw^T}{h_n^2}S(z,\vxw ) f(z|\vxw )dz dP(\vxw ).
	\end{align*}
	By Taylor expansion and \ref{A3}, there exists a $0<\epsilon<1$ such that $S(z,\vxw ) = zS^{(1)}(\epsilon z,\vxw )$.
	We have
	\begin{align*} 
	\E \big\{ h_n^{-2}q_{ni}(\vbeta_n) \big\}&=\int K''\big(\frac{z}{ h_n}+  \widetilde{\vtheta}_n^T \vxw \big) \frac{\vxw \vxw ^T}{h_n^2}zS^{(1)}(\epsilon z,\vxw ) f(z|\vxw )dz dP(\vxw )\\
	&=\int (\zeta-\widetilde{\vtheta}_n^T \vxw )K''(\zeta) \vxw \vxw^T S^{(1)}\big(\epsilon h_n(\zeta-\widetilde{\vtheta}_n^T \vxw ),\vxw \big)  f\big(h_n(\zeta-\widetilde{\vtheta}_n^T \vxw )|\vxw \big)d\zeta dP(\vxw ).
	\end{align*}
	Then by \ref{K1},  \ref{A2},  \ref{A4}-\ref{A5}, and the dominated convergence theorem,
	\begin{align*} 
	\lim\limits_{n \ra \infty}\E \big\{ h_n^{-2}q_{ni}(\vbeta_n) \big\}&=\lim\limits_{n \ra \infty}\int (\zeta-\widetilde{\vtheta}_n^T \vxw )K''(\zeta) \vxw \vxw^T S^{(1)}\big(\epsilon h_n(\zeta-\widetilde{\vtheta}_n^T \vxw ),\vxw \big) f\big(h_n(\zeta-\widetilde{\vtheta}_n^T \vxw )|\vxw \big)d\zeta dP(\vxw )\\
	&=\lim\limits_{n \ra \infty}\int \zeta K''(\zeta) \vxw \vxw^T S^{(1)}(0,\vxw ) f(0|\vxw )d\zeta dP(\vxw )\\
	&=a_2\int \vxw \vxw^T  S^{(1)}(0,\vxw )f(0|\vxw )dP(\vxw ) = \vQ.
	\end{align*}
	
	The Donsker property of $\mathcal{Q}$ implies that for arbitrary $\epsilon>0$,
	$$\sup\limits_{||\vbeta-\vbeta_0||\leq \epsilon h_n}\sup\limits_{h\in(0,1]} \Big|\Big|n^{-1}\sum_{i=1}^{n}\big\{q_{ni}(\vbeta)-\E q_{ni}(\vbeta)\big\}\Big|\Big| = O_p(n^{-1/2}).$$
	Then since $h_n =o(n^{-1/(2b+1)})$ and $(n h_n^4)^{-1}=o(1)$, we can derive that
	\begin{align*}
	\sup\limits_{||\vbeta-\vbeta_0||\leq \epsilon h_n}\Big|\Big|n^{-1} \sum_{i=1}^{n}\big\{h_n^{-2}q_{ni}(\vbeta)-Q\big\}\Big|\Big|&\leq
	\sup\limits_{||\vbeta-\vbeta_0||\leq \epsilon h_n}\Big|\Big|(nh_n^2)^{-1}\sum_{i=1}^{n}\big\{q_{ni}(\vbeta)-\E q_{ni}(\vbeta)\big\}\Big|\Big|+o(1)\\
	&\leq h_n^{-2}\sup\limits_{||\vbeta-\vbeta_0||\leq  \epsilon h_n}\sup\limits_{h\in(0,1]} \Big|\Big|n^{-1}\sum_{i=1}^{n}\big\{q_{ni}(\vbeta)-\E q_{ni}(\vbeta)\big\}\Big|\Big| \\
	&\leq O_p(n^{-1/2}h_n^{-2}) = o_p(1).
	\end{align*}
\end{newproof}

\begin{newproof}{ Lemma~\ref{Hessian}}
	With the consistency result in Theorem \ref{Th1}, we have $P\big(\widehat{\beta}_{n1}=\beta_{01}\big)\ra 1$ as $n\ra \infty$. Hence for any $\vbeta = (\beta_1,\vbetaw^T)^T$ between $\vbeta_0$ and $\vbetah_n$, note that $\indicator(\vx^T\vbeta>0) = \indicator\big(z+\vxw^T(\vbetaw-\vbetaw_0)>0\big)$. Then for $V(\vbeta) =\E_X\{\mu(1,\vx)\indicator(\vx^T\vbeta>0)+\mu(0,\vx)\indicator(\vx^T\vbeta\leq 0)\}$, we have
	\begin{align*}
	V(\vbeta)  &=\E_X\Big\{\mu(1,\vx)\indicator\big(z+\vxw^T(\vbetaw-\vbetaw_0)>0\big) +\mu(0,\vx)\indicator\big(z+\vxw^T(\vbetaw-\vbetaw_0)\leq 0\big)\Big\}\\
	&=  \int\Big[\int_{-\vxw^T(\vbetaw-\vbetaw_0)}^{\infty}\E_X\big\{\mu(1,\vx)|z,\vxw\big\} f(z|\vxw) dz+\int^{-\vxw^T(\vbetaw-\vbetaw_0)}_{-\infty}\E_X\big\{\mu(0,\vx)|z,\vxw\big\} f(z|\vxw) dz\Big]d P(\vxw).
	\end{align*}
	Let $\vdeltaw = \vbetaw-\vbetaw_0$, then $\vdeltaw\rap \widetilde{\vnull}$ for $\vbeta$ between $\vbeta_0$ and $\vbetah_n$, according to Theorem~\ref{Th1}. Note that
	\begin{align*}
	V'(\vbeta)= \frac{\partial V(\vbeta)}{\partial\vbetaw} =& \int S\big(-\vxw^T\vdeltaw, \vxw\big) f\big(-\vxw^T\vdeltaw|\vxw\big) \vxw\vxw^T(\vbetaw_0+\vdeltaw) d P(\vxw),\\
	V''(\vbeta)=& \frac{\partial V'(\vbeta)}{\partial\vbetaw}\\ 
	=& \int  S\big(-\vxw^T\vdeltaw,\vxw\big) \Big\{f\big(-\vxw^T\vdeltaw|\vxw\big)-  (\vxw^T\vbetaw_0)f^{(1)}\big(-\vxw^T\vdeltaw|\vxw\big) \Big\} \vxw\vxw^Td P(\vxw) \\
	&-\int  (\vxw^T\vdeltaw) \Big\{S^{(1)}\big(-\vxw^T\vdeltaw,\vxw\big)f\big(-\vxw^T\vdeltaw|\vxw\big) +S\big(-\vxw^T\vdeltaw,\vxw\big)f^{(1)}\big( -\vxw^T\vdeltaw|\vxw\big)\Big\}\vxw\vxw^Td P(\vxw)\\
	&-\int ( \vxw^T\vbetaw_0)S^{(1)}\big(-\vxw^T\vdeltaw,\vxw\big) f\big( -\vxw^T\vdeltaw|\vxw\big) \vxw\vxw^Td P(\vxw) \\
	=& I_1+I_2 + I_3,
	\end{align*}
	where the definition of $I_i$ ($i=1, 2, 3$) is clear from the context. By Taylor expansion, there exists some constant $0<r_1<1$ such that 
	\begin{align*}
	I_1 =  \int -(\vxw^T\vdeltaw )S^{(1)}\big(-r_1\vxw^T\vdeltaw,\vxw\big) \Big\{f\big(-\vxw^T\vdeltaw|\vxw\big) - (\vxw^T\vbetaw_0)f^{(1)}\big(-\vxw^T\vdeltaw|\vxw\big) \Big\}  \vxw\vxw^Td P(\vxw).
	\end{align*}
	By \ref{A2}, \ref{A4}-\ref{A5}, we know that  the components of $\vxw $, $S^{(i)}(z,\vxw )$ and $f^{(i)}(z|\vxw )$, $i=0,1$, are bounded for almost every $\vxw $. Then for any $\vdeltaw\rap \widetilde{\vnull}$, it is easy to conclude $I_1\rap 0$ and $I_2\rap 0$. To evaluate $I_2$, note that for some constant $0<r_2<1$,
	\begin{align*}
	I_2 =  \vI_V +\int   (\vdeltaw^T\vxw\vxw^T\vbetaw_0)\Big\{&S^{(2)}\big(-r_2\vxw^T\vdeltaw,\vxw\big) f\big(-\vxw^T\vdeltaw|\vxw\big)\\ +& S^{(1)}\big(-\vxw^T\vdeltaw,\vxw\big) f^{(1)}\big(-r_2\vxw^T\vdeltaw|\vxw\big)\Big\} \vxw\vxw^Td P(\vxw).
	\end{align*}
	With the boundedness of the components of $\vxw$, $S^{(1)}(z,\vxw)$, $S^{(2)}(z,\vxw)$, $f(z|\vxw)$ and $f^{(1)}(z|\vxw)$ from \ref{A2}, \ref{A4}-\ref{A5}, we also have $I_2 \rap \vI_V$ as $\vdeltaw\rap \widetilde{\vnull}$, where $\vI_V$ is negative definite by \ref{A5}. This finishes the proof.
\end{newproof}

Recall from Section~\ref{Theory2} of the main paper that
$r=\{r_1, \ldots, r_n\}$ denotes the collection of the random bootstrap weights and
$w=\{W_1, \ldots, W_n\}$ denotes the random sample of observations, where
$W_i=(\vx_i, A_i, Y_i)$.  Given a sequence of random variables $R_n$, $n=1, \ldots, n$,
we write $R_n=\opr$ if for any $\epsilon>0, \delta>0$,  we have $P_{w}\big(P_{r|w}(|R_n| >\epsilon)>\delta\big)\ra 0$ as $n\ra \infty$. 
In the bootstrap literature,  $R_n$ is said to converge to zero in probability, conditional on the data.
Let $\E_{r|w}$ and $\Var_{r|w}$ denote the conditional expectation and the conditional variance according to the
distribution of $r$ given $x$.
Furthermore, $\oprw$ denotes a random sequence that converges to zero in probability with respect to the joint distribution of $(r, w)$,
and  $\opw$ 
denotes a random sequence that converges to zero in probability with respect to the distribution of $r$ only.
By Lemma 3 of  \citet{ChengHuang}, if
$R_n=o_{p_{rw}}(1)$, then  $R_n=o_{p_{r}}(1)$. In particular, if  $R_n$ depends only on the data $w$ but not
on the random weights $r$ and if $R_n=o_{p_{w}}(1)$, then
it is easy to see $R_n=o_{p_{rw}}(1)$, and hence it is $o_{p_{r}}(1)$. In this part of proof, 
we will include subscripts in the probability and expectation to clarify which probability distribution
is used in the calculation.
\vspace{1em}

\begin{newproof}{ Lemma~\ref{L78}}
	It suffices to prove\\
	(a) let  $\vtheta_n^* = (\vbetah_n^*-\vbetah_n)/h_n$,  we have $\vtheta_n^*  = \opr$;\\
	(b) let $\{\vbeta_n\} = \{(\beta_{n1}, \vbetaw_n^T)^T\}$ be any sequence in $\bbB$ such that $(\vbeta_n-\vbetah_n)/h_n\ra 0$ as $n\ra\infty$, then $\vQ_n^*(\vbeta_n; h_n) = \vQ + \opr$.
	
	To prove (a), 	for any $\vtheta \in \bbR^{p-1}$, define  
	$\vR_n^*(\vtheta) = \frac{2}{nh_n^2}\sum_{i=1}^nr_i(2A_i-1)K'\big(\frac{z_i}{ h_n}+  \vtheta^T \vxw _i\big) \vxw _iY_i$. We observe 
	\begin{align*} 
	\big|\big|\E _{r|w}\vR^*_n(\vtheta) - \vQ\vtheta\big|\big|& \leq \big|\big|\E_{r|w} \vR^*_n(\vtheta) - \E_w  \vR_n(\vtheta)\big|\big| +\big|\big|\E_w  \vR_n(\vtheta) - \vQ\vtheta\big|\big| \\
	&= \big|\big|R_n(\vtheta) - \E_w  \vR_n(\vtheta)\big|\big| +\big|\big|\E_w \vR_n(\vtheta) - \vQ\vtheta\big|\big|.
	\end{align*} 
	By Lemma~\ref{L3},  
	\begin{align*} 
	\sup\limits_{\vtheta\in\Theta_n}\big|\big|\vR_n(\vtheta)-&\E_w  \vR_n(\vtheta)\big|\big| = o_p (1) \pxp,\\
	\big|\big|\E_w  \vR_n(\vtheta) - \vQ\vtheta\big|\big| \leq o&(1)+\alpha_1h_n||\vtheta|| +\alpha_2h_n||\vtheta||^2,
	\end{align*}
	uniformly over  $\vtheta\in\Theta_n \pxp$ for some  finite $\alpha_1$ and $\alpha_2$. Hence
	$$\big|\big|\E _{r|w}\vR^*_n(\vtheta) - \vQ\vtheta\big|\big| \leq o(1)+\alpha_1h_n||\vtheta|| +\alpha_2h_n||\vtheta||^2,$$
	uniformly over  $\vtheta\in\Theta_n\pxp$.
	By Theorem~\ref{Th3}, $h_n\vtheta_n^* =  \opr$. So $\vR_n^*(\vtheta^*_n) =\opr$. So we have
	$$\big|\big|\vQ \vtheta^*_n\big|\big|\leq o(1)+\alpha_1h_n||\vtheta_n^*|| +\alpha_2h_n||\vtheta_n^*||^2.$$
	Then similarly to the proof of Lemma B2, we can show that $\vtheta_n^*  = \opr$.
	
	To prove (b), let $q^*_{ni}(\vbeta)= (2A_i-1)rK''\big(\frac{\vx_i^T\vbeta}{h}\big)\vxw_i\vxw_i^T Y_i$. It suffices to show that $$\sup\limits_{||\vbeta-\vbetah_n||\leq \epsilon h_n}\Big|\Big|n^{-1}\sum_{i=1}^{n}\big\{h_n^{-2}q^*_{ni}(\vbeta)- Q\big\}\Big|\Big|=\oprw,$$
	for arbitrary positive $\epsilon$.
	
	Let $\vbetaw_n^r=\vbetaw_n+h_n\widetilde{\vtheta}_n^*$ with $\widetilde{\vtheta}_n^*\ra0$. Consequently, $\lim\limits_{n\ra\infty}\E_{w}\E_{r|w} \{h_n^{-2}q^*_{ni}(\vbeta_n^r)\}=Q$.
	
	The Donsker property of $\mathcal{Q}^*$ implies that for arbitrary $\epsilon>0$,
	$$\sup\limits_{||\vbeta-\vbeta_n||\leq \epsilon h_n}\sup\limits_{h\in(0,1]} \Big|\Big|n^{-1}\sum_{i=1}^{n}\big\{q^*_{ni}(\vbeta)-\E_{w}\E_{r|w}  q^*_{ni}(\vbeta)\big\}\Big|\Big| = O_{p_{rw}}(n^{-1/2}).$$
	Then since $h_n =o(n^{-1/(2b+1)})$ and $(n h_n^4)^{-1}=o(1)$, we can derive that
	\begin{align*}
	\sup\limits_{||\vbeta-\vbeta_n||\leq \epsilon h_n}n^{-1}\Big|\Big|\sum_{i=1}^{n} \big\{h_n^{-2}q^*_{ni}(\vbeta)-Q\big\} \Big|\Big|&\leq
	\sup\limits_{||\vbeta-\vbeta_n||\leq \epsilon h_n}(nh_n^2)^{-1}\Big|\Big|\sum_{i=1}^{n} \big\{q^*_{ni}(\vbeta) -\E_{w}\E_{r|w}  q^*_{ni}(\vbeta)\big\}\Big|\Big|+o(1)\\
	&\leq h_n^{-2}\sup\limits_{||\vbeta-\vbeta_n||\leq \epsilon h_n}\sup\limits_{h\in(0,1]} \Big|\Big|n^{-1}\sum_{i=1}^{n} \big\{q^*_{ni}(\vbeta) -\E_{w}\E_{r|w} q^*_{ni}(\vbeta)\big\}\Big|\Big|\\
	&\leq O_{p_{rw}}(n^{-1/2}h_n^{-2}) =\oprw.
	\end{align*}
\end{newproof}

\section{Moving Parameter Asymptotics} \label{moving}
To better understand the behavior of the proposed inference procedure, we study the properties of the smoothed estimator and its bootstrapped version under a moving parameter or local asymptotic framework, as motivated by \cite{laber2010}.

Consider the following semiparametric model
\bqan\label{local}
Y = \mu(\vx) +\vx^T(\vbeta_0+b_n\vs)\vx A+\epsilon,
\eqan 
where $\mu(\vx)$ is an unspecified function, $\epsilon$ is a sub-Gaussian random error term with mean zero and variance $\sigma^2$. The local model (\ref{local}) perturbs $\vbeta_0 = (\beta_{01},\vbetaw_0)$ (with $|\beta_{01}| = 1$) by a small quantity $b_n\vs$, with $b_n$ being a sequence of real numbers that converges to zero as $n\ra \infty$ and $\vs = (s_1,\vsw)$ is a fixed $p$-dimensional vector.
We write $\vs = (s_1,\vsw)$ and assume $s_1=0$ to avoid complications that are not relevant to the main results. When $b_n=0$, the optimal treatment regime is given by $I(\vx^T\vbeta_0>0)$.

Consider a random sample $\{(\vx_i,A_i,Y_i), i=1,...,n\}$ from (\ref{local}).
We estimate $\vbeta_0$ by the smooth robust estimator introduced in Section \ref{dragon}, that is, $\vbetah_n =\arg\max_{ \vbeta\in \bbB} n^{-1}\sum_{i=1}^n(2A_i-1) K\big(\frac{\vx_i^T\vbeta}{h_n}\big)Y_i$.
Correspondingly, the confidence interval is constructed using the formula in (\ref{b1}) based on the bootstrapped estimator 
$\vbetah_n^{*} =\arg\max_{ \vbeta\in \bbB} n^{-1}\sum_{i=1}^nr_i(2A_i-1) K\big(\frac{\vx_i^T\vbeta}{h_n}\big)Y_i$.
That is, we study the behavior of the procedures proposed earlier which are constructed in a model-free fashion when the underlying data are generated by (\ref{local}).
To study the local asymptotics, define 
$$\vD_0 = 2a_1\E\big[\vxw \vxw^T f(0|\vxw )\big\{\E(\mu^2(\vx)|z=0,\ \vxw ) + \sigma^2\big\}\big]\mbox{\quad and \quad }\vQ_0 = a_2\E\big\{\vxw \vxw^T f(0|\vxw )\big\},$$
where $a_i$ ($i=1$, 2) is defined in Section \ref{Theory1}.
As before, write $\vbetah_n = (\vbetah_{n1},\vbetaw_n^T)^T\in\bbR^p$ and $\vbetah^*_n=(\widehat{\beta}^*_{n1}, \widetilde{\vbeta}^{*T}_n)^T$.

The following two theorems show that asymptotic normality holds for $\vbetah_n$ and $\widetilde{\vbeta}^{*T}_n$ for $b_n$ chosen at appropriate rate. If the sequence $b_n$ goes to zero faster that $(nh_n)^{-1/2}$, the smoothed estimator is asymptotically unbiased and the bootstrap confidence interval for $\vbeta_0$ is asymptotically accurate.  The proofs of these results can be found in Section~\ref{Movingproof}.  
\setcounter{theorem}{4} 
\begin{theorem}
	\label{Th5}
	Assume $K(\cdot)$ satisfies \ref{K1} -  \ref{K3} for some $b\geq 2$,
	$h_n=o(n^{-1/(2b+1)})$ and  $n^{-1}h_n^{-4}=o(1)$.  If $b_n = (nh_n)^{-1/2}$, then under  \ref{A1},  \ref{A2},  \ref{A4}, 	
	\bqa
	\sqrt{nh_n}(\vbetaw_n-\vbetaw_0) \ra N\big(-a_2^{-1}\vsw, \vQ_0^{-1}\vD_0\vQ_0^{-1}\big)
	\eqa
	in distribution as $n\ra \infty$.
\end{theorem}
\begin{theorem}
	\label{Th6}
	Assume $K(\cdot)$ satisfies \ref{K1} -  \ref{K3}, for some $b\geq 2$,
	$h_n=o(n^{-1/(2b+1)})$, and  $\log(n)=o(nh_n^4)$. If $b_n = (nh_n)^{-1/2}$, then under  \ref{A1},  \ref{A2},  \ref{A4}, \ref{A6}, 
	\bqa 	
	\sqrt{nh_n}(\vbetaw_n^{*}-\vbetaw_n)= N\big(-a_2^{-1}\vsw, \vQ_0^{-1}\vD_0\vQ_0^{-1}\big)+\opr.
	\eqa
\end{theorem}

\section{Proof of  Results in Section~\ref{moving}}  \label{Movingproof}
Let $Y$ be the response generated from the local model (\ref{local}). We can write 
$Y = \widecheck{Y} +  b_n\widetilde{Y}$, where $\widecheck{Y}= \mu(\vx) +A\vx^T \vbeta_0+\epsilon$ and $\widetilde{Y} = A\vxw^T\vsw$.
Note that $\widecheck{Y}$ satisfied all the assumptions about the outcome variable in \ref{A1}, \ref{A3} and \ref{A5}. It follows that all the preceding lemmas and theorems still hold if regarding $\widecheck{Y}$ as the observed response $Y$. In addition, since $(2A-1)\widetilde{Y}$ is square integrable and does not depend on $(\vbeta,h)$, it implies that all classes listed in Lemma~\ref{Moon} are still VC classes with $\widetilde{Y}$ as their responses. In the following proof, we use ``$\ \widecheck{\ }\ $'' to denote corresponding notation when we replace $Y$ with $\widecheck{Y}$.
For example, we define
$$\widetilde{\widecheck{M}}_n(\vbeta, h_n) = 2n^{-1}\sum_{i=1}^n(2A_i-1)\indicator(\vx_i^T\vbeta> 0)\widecheck{Y}.$$

First we will prove the consistency of the smoothed estimator given the observed data 
$\{(\vx_i,A_i,Y_i), i=1,...,n\}$ from (\ref{local}).
\begin{lemma}
	\label{moving_consistency}
	Under (A1), (A2) and assume $K(\cdot)$ satisfies (K1), if $b_n = o(1)$, then  $\vbetah_n = \vbeta_0+o_p(1)$.
\end{lemma}

\begin{newproof}{ Lemma~\ref{moving_consistency}}
	We observe that $\vbetah_n$ maximizes $\widetilde{M}_n(\vbeta, h_n)$ over $\vbeta\in \bbB$, and $\vbeta_0$ maximizes $\widecheck{M}(\vbeta)$. Note that
	\begin{align*}
	\sup\limits_{\vbeta  \in \bbB} \big|\widetilde{M}_n(\vbeta, h_n) -  \widecheck{M}(\vbeta)\big| \leq \sup\limits_{\vbeta  \in \bbB} \big|2b_nn^{-1}\sum_{i=1}^n (2A_i-1) K\big(\frac{\vx_i^T\vbeta}{h_n}\big)\widetilde{Y}_i\big| + \sup\limits_{\vbeta  \in \bbB} \big|\widetilde{\widecheck{M}}_n(\vbeta, h_n) -  \widecheck{M}(\vbeta)\big| .
	\end{align*}
	
	Lemma~\ref{Lemma5} implies that $\sup\limits_{\vbeta  \in \bbB} \big|\widetilde{\widecheck{M}}_n(\vbeta, h_n) -  \widecheck{M}(\vbeta)\big|=o_p(1)$.
	In addition, it is obvious that
	\begin{align*}
	\sup\limits_{\vbeta  \in \bbB} \big|\frac{b_n}{n}\sum_{i=1}^n (2A_i-1) K\big(\frac{\vx_i^T\vbeta}{h_n}\big)\widetilde{Y}_i\  \big| \leq & \sup\limits_{\vbeta  \in \bbB} \Big|\frac{b_n}{n}\sum_{i=1}^n\Big[ (2A_i-1)K\big(\frac{\vx_i^T\vbeta}{h_n}\big)\widetilde{Y}_i- \E \big\{ K\big(\frac{\vx_i^T\vbeta}{h_n}\big)\vxw_i^T\vsw\big\}\Big]\Big|\\
	&+ \sup\limits_{\vbeta  \in \bbB} \big|b_n\E  \big\{K\big(\frac{\vx_i^T\vbeta}{h_n}\big)\vxw_i^T\vsw\big\}\big|.
	\end{align*}
	The Donsker property of $\mathcal{F}$ ensures the first term converges to 0 in probability if $b_n = o(\sqrt{n})$. By the boundedness of $K(\cdot)$ and $\vx$, the second term also goes to 0 as $b_n = o(1)$. So  $\sup\limits_{\vbeta  \in \bbB} \big|\widetilde{M}_n(\vbeta, h_n) -  \widecheck{M}(\vbeta)\big| =o_p(1)$ can be concluded.
	
	The construction of $\widecheck{Y}$  implies that for every $\tau>0$, 
	\begin{align*}
	\sup\limits_{||\vbeta-\vbeta_0|| > \tau} \widecheck{M}(\vbeta)-\widecheck{M}(\vbeta_0) &=\sup\limits_{\vbeta  \in \bbB}  2\E [(2A_i-1)\widecheck{Y}_i \{I(\vx_i^T\vbeta>0)-I(\vx_i^T\vbeta_0>0)\}]\\
	&=\sup\limits_{||\vbeta-\vbeta_0|| > \tau} \E [\vx_i^T\vbeta_0 \{(\vx_i^T\vbeta>0)-I(\vx_i^T\vbeta_0>0)\} ]< 0.
	\end{align*} 
	Hence, $\vbetah_n = \vbeta_0+o_p(1)$ is derived from Theorem 5.7 in van der Vaart (2000).
\end{newproof}


\begin{newproof}{ Theorem~\ref{Th5}}
	The proof of Theorem~\ref{Th2} implies that it suffices to verify: \\
	(a) $(nh_n)^{1/2}\vT_n(\vbeta_0; h_n) \rad N(a_2^{-1}\vQ_0\vsw, \vD_0)$;\\
	(b) $\vQ_n(\vbeta_n^r; h_n) = \vQ_0+o_p(1)$ for any $\vbeta_{n}^r$ is between $\vbetah_n$ and $\vbeta_0$.
	
	
	To prove (a), note that Lemma~\ref{L1} indicates that 
	$$\E (nh_n)^{1/2}\big\{ \vT_n (\vbeta_0; h_n) \big\} \ra a_2^{-1}\vQ_0\vsw\mbox{, \quad and \quad}\Var(nh_n)^{1/2}\big\{\vT_n(\vbeta_0; h_n) \big\} \ra \vD_0.$$ 
	It is sufficient to prove that 
	$(nh_n)^{1/2}\vgamma^T\big\{\vT_n(\vbeta_0; h_n)-\E \vT_n(\vbeta_0; h_n)\big\}$ is asymptotically 
	$N(\widetilde{\vnull},\vgamma^T\vD_0\vgamma)$ for any fixed vector $\vgamma\in\bbR^{p-1}$ such that $||\vgamma||=1$. Define 
	\begin{align*}
	q_i &= 2(2A_i-1)(nh_n)^{1/2}K'\big(\frac{\vx_i^T\vbeta_0}{ h_n} \big) \frac{\vgamma^T\vxw _i}{h_n}Y_i \\
	&= \widecheck{q}_i + 2(2A_i-1)K'\big(\frac{\vx_i^T\vbeta_0}{ h_n} \big) \frac{\vgamma^T\vxw _i}{h_n}\widetilde{Y}_i\\
	&= \widecheck{q}_i +\widetilde{q}_i 
	\end{align*}
	for $ \widecheck{q}_i $ defined as in the proof of Theorem~\ref{Th2}. With Lyapunov central limit theorem, we will verify 
	\begin{align}
	\lim\limits_{n \ra \infty} (s_n)^{-4} \sum_{i=1}^n \E \big\{(q_i-\E q_i)^4\big\}=0,
	\label{dolphin_moving}
	\end{align}
	where
	$\lim\limits_{n \ra \infty}(n^{-1}s_n)^2= \lim\limits_{n \ra \infty}\sum_{i=1}^n \Var(n^{-1}q_i)= \vgamma^T\vD_0\vgamma$. The fact that $\lim\limits_{n \ra \infty} n^{-3} \E \big\{(\widecheck{q}_i -\E \widecheck{q}_i )^4\big\}=0$ implies that the left-side of (\ref{dolphin_moving}) is bounded from above (up to a positive constant) by
	\begin{align*}
	\lim\limits_{n \ra \infty} n^{-3} \E (\widetilde{q}_i^4) +\lim\limits_{n \ra \infty}n^{-3}(\E \widetilde{q}_i )^4 =&\lim\limits_{n \ra \infty} (n^3h_n^4)^{-1}8\E \big\{K'(\frac{x_i^T\vbeta_0}{ h_n})^4( \vgamma^T\vxw _i\vxw _i^T\vsw)^4\big\}\\
	&+\lim\limits_{n \ra \infty} (n^3h_n^4)^{-1}\big[\E \big\{K'(\frac{x_i^T\vbeta_0}{ h_n})( \vgamma^T\vxw _i\vxw _i^T\vsw)\big\}\big]^4.
	\end{align*}
	With the boundedness of $K'(\cdot)$ and $\vxw$, $(n^3h_n^4)^{-1} = o(1)$ implies the Lyapunov condition is satisfied, and (a) follows. To prove (b), note that
	\begin{align*}
	\sup\limits_{\vbeta\in\bbB}||\vQ_n(\vbeta; h_n)- \vQ_n(\vbeta; h_n)||& = \sup\limits_{\vbeta\in\bbB}\Big|\Big|\frac{2b_n}{n}\sum_{i=1}^n (2A_i-1) K''\big(\frac{\vx_i^T\vbeta}{ h_n} \big) \frac{\vxw_i\vxw_i^T}{h_n^2}\widetilde{Y}_i \Big|\Big|\\
	&\leq O_p\big((n^2h_n^5)^{-1/2}\big) + O(b_nh_n^{-1})\sup\limits_{\vbeta\in\bbB}\Big|\Big|\E \big\{K''\big(\frac{\vx_i^T\vbeta}{ h_n} \big) \frac{\vxw_i\vxw_i^T}{h_n}\vxw^T\vsw\big\}\Big|\Big|\\
	&= O_p\big((n^2h_n^5)^{-1/2}\big) + O\big((nh_n^{3})^{-1/2}\big) = o_p(1),
	\end{align*}
	since $\mathcal{Q}$ is a VC class. Then it suffices to show that $\vtheta_n = (\vbetah_n-\vbeta_0 )/h_n = o_p(1)$. Consider $\vR_n(\vtheta)$ defined as in Lemma~\ref{L3}.
	The Donsker properties of $\mathcal{H}$ imply that 
	\begin{align*}
	\sup\limits_{\vtheta\in\Theta_n}||\vR_n(\vtheta)-\widecheck{\vR}_n(\vtheta)||& \leq O_p\big((n^2h_n^5)^{-1/2}\big) + O(b_nh_n^{-1})\sup\limits_{\vtheta\in\Theta_n}\Big|\Big|\E \big\{K'\big(\frac{z_i}{ h_n}+  \vtheta^T \vxw _i\big) \frac{\vxw _i\vxw_i^T\vsw}{h_n}\big\}\Big|\Big|\\
	&=O_p\big((n^2h_n^5)^{-1/2}\big) + O\big((nh_n^{3})^{-1/2}\big) = o_p(1),
	\end{align*}
	where $\Theta_n$ is defined in Lemma A3. Combined with Lemma A3, it implies that $\sup\limits_{\vtheta\in\Theta_n}||\vR_n(\vtheta)- \vQ_0\vtheta|| \leq o(1)+\alpha_1h_n||\vtheta|| +\alpha_2h_n||\vtheta||^2$. By the definition of $\vtheta_n$, we know that $h_n\vtheta_n\rap 0$, and $\vR_n(\vtheta_n)\rap 0$. Then from the proof of Lemma~\ref{L45}, (b) can be concluded. 
\end{newproof}

For the asymptotic distribution for bootstrap estimators with the moving parameter framework, we first prove its consistency. 
\begin{lemma}
	\label{moving_boots_cons}
	Under (A1), (A2), (A6) and assume $K(\cdot)$ satisfies (K1), if $b_n = o(1)$, then $\vbetah^{*}_n = \vbetah_n +\opr$.
\end{lemma}

\begin{newproof}{ Lemma~\ref{moving_boots_cons}}
	By definition, $\vbetah^{*}_n$ maximizes $\widetilde{M}^{*}_n(\vbeta, h_n)$ over $\vbeta\in \bbB$. 
	First, given $\mathcal{F}^{*new}$ is a VC class, the Donsker property and Lemma~\ref{Lemma5} jointly indicate that
	\begin{align*}
	\sup\limits_{\vbeta  \in \bbB}\big|\widetilde{M}^{*}_n(\vbeta, h_n) -\widetilde{M}_n(\vbeta, h_n) \big| \leq& \sup\limits_{\vbeta  \in \bbB}\Big|\frac{2b_n}{n}\sum_{i=1}^n(r_i-1)(2A_i-1) K\big(\frac{\vx_i^T\vbeta}{h_n})\widetilde{Y}_i \Big|\\
	&+\sup\limits_{\vbeta  \in \bbB}\big|\widetilde{\widecheck{M} }^{*}_n(\vbeta, h_n) -\widetilde{\widecheck{M} }_n(\vbeta, h_n) \big|= \oprw.
	\end{align*}
	By Lemma~3 of Cheng \& Huang (2010),
	$\sup\limits_{\vbeta  \in \bbB}  \big|\widetilde{M}^{*}_n(\vbeta, h_n) -\widetilde{M}_n(\vbeta, h_n)  \big|  =\opr$.
	By Theorem 5.7 in van der Vaart (2000), to prove the theorem, it is sufficient to show that for any $\epsilon > 0$,
	\bqan\label{bear1}
	\lim\limits_{n\ra\infty}P_{w}\Big(\sup\limits_{||\vbeta-\vbetah_n||> \epsilon} 
	\big\{ \widetilde{M}_n(\vbeta, h_n) - \widetilde{M}_n(\vbetah_n, h_n) \big\}<0\Big)= 1.
	\eqan 
	Note that Lemma~\ref{Lemma5} and the consistency of $\vbetah_n$ implies that 
	\begin{align*}
	\sup\limits_{||\vbeta-\vbetah_n||> \epsilon} 
	\big\{ \widetilde{M}_n(\vbeta, h_n) - \widetilde{M}_n(\vbetah_n, h_n) \big\} \leq& \sup\limits_{||\vbeta-\vbetah_n||> \epsilon} 
	\big\{ \widetilde{\widecheck{M}}_n(\vbeta, h_n) - \widetilde{\widecheck{M}}_n(\vbeta_0, h_n)\big\} +\big\{ \widetilde{\widecheck{M}}_n(\vbeta_0, h_n) - \widetilde{\widecheck{M}}_n(\vbetah_n, h_n)\big\} \\
	&+ \sup\limits_{||\vbeta-\vbetah_n||> \epsilon} \Big| \frac{2b_n}{n} \sum_{i=1}^n(2A_i-1) \Big\{K\big(\frac{\vx_i^T\vbeta}{h_n}\big)- K\big(\frac{\vx_i^T\vbetah_n}{h_n}\big)\Big\}\widetilde{Y}_i \Big|\\
	=&\sup\limits_{||\vbeta-\vbetah_n||> \epsilon} 
	\big\{\widecheck{M}(\vbeta) - \widecheck{M}(\vbeta_0) \big\} + o_p(1).
	\end{align*}
	Furthermore, the consistency of $\vbetah_n$ implies that for all sufficiently large 
	$n$, any $\vbeta$ that satisfies $||\vbeta- \vbetah_n||> \tau$ would also satisfy $||\vbeta-\vbeta_0||\geq \tau/2$.   Hence, Lemma~\ref{moving_consistency} implies (\ref{bear1}) holds.	
\end{newproof}


\begin{newproof}{ Theorem~\ref{Th6}}
	The proofs of Theorem~\ref{Th4} indicate that it is sufficient to verify: \\
	(a) $\vQ^{*}_n(\vbeta_n^{*r}; h_n) = \vQ_0+\opr$ for any $\vbeta_{n}^{*r}$ is between $\vbetah^{*}_n$ and $\vbetah_n$;\\
	(b) $(nh_n)^{1/2}\vT_n^{*}(\vbetah_n; h_n) = N(a_2^{-1}\vQ_0\vsw, \vD_0)+\opr$.
	
	To prove (a), the fact that $\mathcal{Q}^{*}$ is a VC class implies that
	\begin{align*}
	\sup\limits_{\vbeta\in\bbB}\big|\big|\vQ^{*}_n(\vbeta; h_n)- \widecheck{\vQ}_n^{*}(\vbeta; h_n)\big|\big|& = \sup\limits_{\vbeta\in\bbB}\Big|\Big|\frac{2b_n}{n}\sum_{i=1}^n r_i(2A_i-1) K''\big(\frac{\vx_i^T\vbeta}{ h_n} \big) \frac{\vxw_i\vxw_i^T}{h_n^2}\widetilde{Y}_i\Big|\Big|\\
	&\leq O_p\big((n^2h_n^5)^{-1/2}\big) + O(b_nh_n^{-1})\sup\limits_{\vbeta\in\bbB}\Big|\Big|\E_w \big\{K''\big(\frac{\vx_i^T\vbeta}{ h_n} \big) \frac{\vxw_i\vxw_i^T}{h_n}\vxw^T\vsw\big\}\Big|\Big|\\
	&= O_p\big((n^2h_n^5)^{-1/2}\big) + O\big((nh_n^{3})^{-1/2}\big) = \opr.
	\end{align*}	
	It suffices to show that $\vtheta_n^{*} = (\vbetah^{*}_n-\vbeta_0 )/h_n = \opr$. For $\vR_n^{*}(\vtheta) $ defined in the proof of Lemma~\ref{L78}, the fact that $\mathcal{H}^{*}$ is a VC class indicates that 
	\begin{align*}
	\sup\limits_{\vtheta\in\Theta_n}||\vR^{*}_n(\vtheta)-\widecheck{\vR}_n^{*}(\vtheta)|| &\leq O_p\big((n^2h_n^5)^{-1/2}\big) + O(b_nh_n^{-1})\sup\limits_{\vtheta\in\Theta_n}\Big|\Big|\E_w \big\{K'\big(\frac{z_i}{ h_n}+  \vtheta^T \vxw _i\big) \frac{\vxw _i\vxw_i^T\vsw}{h_n}\big\}\Big|\Big|\\
	&=O_p\big((n^2h_n^5)^{-1/2}\big) + O\big((nh_n^{3})^{-1/2}\big) = \opr,
	\end{align*}
	where $\Theta_n$ is defined in Lemma A3. Combined with Lemma B4, it implies that $\sup\limits_{\vtheta\in\Theta_n}||\vR^{*}_n(\vtheta)- \vQ_0\vtheta|| \leq o(1)+\alpha_1h_n||\vtheta|| +\alpha_2h_n||\vtheta||^2$. By the definition of $\vtheta_n^{*}$, we know that $h_n\vtheta_n^{*}=\opr$, and $\vR_n^{*}(\vtheta_n^{*})=\opr$. Then from the proof of Lemma B4, (a) can be concluded. To prove (b),  the proof of Lemma A5 and the VC class $\mathcal{H}^{*}$ implies that $(nh_n)^{1/2}\{\vT_n^{*}(\vbetah_n^{}; h_n) - \vT_n^{*}(\vbeta_0; h_n) \}= \opr$. Then observe that
	\begin{align*} 
	\lim\limits_{n\ra \infty}\E_w\E_{r|w}  \big\{(&nh_n)^{1/2}\vT_n^{*}(\vbeta_0; h_n) \big\}= a_2^{-1}\vQ_0 \vsw, \\
	\lim\limits_{n\ra \infty}\E_w\Big[\Var_{r|w}& \big\{(nh_n)^{1/2}\vT_n^{*}(\vbeta_0; h_n)\big\}  \Big]= \vD_0.
	\end{align*}
	It is sufficient to prove that 
	$(nh_n)^{1/2}\vgamma^T\big\{\vT^{*}_n(\vbeta_0; h_n)-\E \vT^{*}_n(\vbeta_0; h_n)\big\}=N(\widetilde{\vnull},\vgamma^T\vD_0\vgamma)+\opr$ for any fixed vector $\vgamma\in\bbR^{p-1}$ such that $||\vgamma||=1$. Define 	
	\begin{align*}
	q_i ^{*}&= 2r_i(2A_i-1)(nh_n)^{1/2}K'\big(\frac{\vx_i^T\vbeta_0}{ h_n} \big) \frac{\vgamma^T\vxw _i}{h_n}Y_i \\
	&= \widecheck{q}_i^{*} + 2r_i(2A_i-1)K'\big(\frac{\vx_i^T\vbeta_0}{ h_n} \big) \frac{\vgamma^T\vxw _i}{h_n} \widetilde{Y}_i= \widecheck{q}_i^{*}+\widetilde{q}_i^*,
	\end{align*}
	for $\widecheck{q}^{*}_i$ defined as in the proof of Theorem 4. To check the Lyapunov condition, it suffices to prove 
	\begin{align*}
	\lim\limits_{n \ra \infty} (s_n^{*})^{-4} \sum_{i=1}^n \E \big\{(q_i^{*}-\E q_i^{*})^4\big\}=0,
	\end{align*}
	where $(s_n^{*} )^2= \sum_{i=1}^n\Var_{r|w}(q_i^{*}) $. Similarly as the proof of Theorem~\ref{Th5}, the Lyapunov condition holds if
	$$(s_n^{*})^{-4}  \sum_{i=1}^n \E_{r|w} (\widetilde{q}_i^{*4})\xrightarrow{a.s.} 0, \quad \mbox{and} \quad
	(s_n^{*})^{-4}  \sum_{i=1}^n  (\E_{r|w}\widetilde{q}_i^*)^4 \xrightarrow{a.s.} 0. $$
	
	Since $r$ is sub-Gaussian, then $\E|r|^k$ is finite  for any positive integer $k$. Hence with bounded $K(\cdot)$ and $\vxw$ and fixed $\vs$,  the strong law of large numbers and the continuous mapping theorem imply that
	\begin{align*}
	(s_n^{*})^{-4}  \sum_{i=1}^n \E_{r|w} (\widetilde{q}_i^{*4} )\xrightarrow{a.s.} \{\lim\limits_{n\ra \infty}\E_{w}(s_n^{*})^{2}\}^{-2}\lim\limits_{n\ra \infty}\E_{w} \sum_{i=1}^n\E_{r|w}(\widetilde{q}_i^{*4}) =0,\\
	(s_n^{*})^{-4}  \sum_{i=1}^n (\E_{r|w} \widetilde{q}_i^* )^4\xrightarrow{a.s.} \{\lim\limits_{n\ra \infty}\E_{w}(s_n^{*})^{2}\}^{-2}\lim\limits_{n\ra \infty}\E_{w} \sum_{i=1}^n(\E_{r|w}\widetilde{q}_i^*)^4 =0.
	\end{align*}
	This verifies the Lyapunov condition and (b) follows. 
\end{newproof}

\newpage
\section{Pseudo Codes for the Proximal Algorithm} \label{Algo}
\begin{algorithm} [!ht]
	\caption{Proximal ($\vbeta^{(0)}$, $\alpha_0$, $\gamma$)} \label{Proximal} 
	\begin{algorithmic}[1]
		\State Set $t=0$.
		\State Set diff = 0.
		\While {diff $\geq 0$} 
		\State  $t \leftarrow t+1$.
		\State $\alpha_t \leftarrow \gamma\alpha_{t-1}$.
		\State $\delta_t\leftarrow (n\alpha_t)^{-1}\sum_{i=1}^n(2A_i-1) K'\big(\frac{\vx_i^T\beta^{(t-1)}}{h_n}\big)\frac{\vx_i}{h_n}Y_i$. 
		\State  $\vbeta^{(t)}  \leftarrow  \beta^{(t-1)} + \delta_t$. 
		\State diff $ \leftarrow  \frac{2}{n}\sum_{i=1}^n(2A_i-1)Y_i\big\{K\big(\frac{\vx_i^T\vbeta^{(t)}}{h_n}\big)-K\big(\frac{\vx_i^T\vbeta^{(t-1)}}{h_n}\big)-h_n^{-1}\vx_i^T\delta_tK'\big(\frac{\vx_i^T\vbeta^{(t-1)}}{h_n}\big)\big\} +\alpha_t||\delta_t||^2.$
		\EndWhile	
		\State Output $\vbeta^{(t)}$.	
	\end{algorithmic} 
\end{algorithm}

\section{Additional numerical results}\label{numeric}
\setcounter{table}{0}
\renewcommand{\thetable}{S\arabic{table}} 

\noindent {\bf Example 1} (binary response). 
The binary response $Y$ is generated to satisfy $\E Y =  ( 1+e^{-\vx^T\vbeta^{opt}})^{-1}$, where
$\vX$, $\vbeta^{opt}$, $A$ are the same as in Settings 1 \& 2 in the main paper. 
Table~\ref{table: Binary} and Table~\ref{table: binary_inf} summarize the simulations results. 
We observe satisfactory performance as in the continuous response cases.
\begin{table}[!ht]
	\centering
	\def\~{\hphantom{0}}
	\caption{Monte Carlo estimates of 
		the bias and standard deviation of the estimate for the parameters indexing the optimal treatment regime, the match ratio (percentage of times the estimated optimal treatment regime matches the theoretically optimal treatment regime), and the bias and standard deviation of the estimated optimal value with binary outcomes.}
	\label{table: Binary}
	\begin{tabular} {ccccccc} 
		\hline 
		$n$ &$\beta^{opt}_0$&$\beta^{opt}_1$  &  $\beta^{opt}_2$  &  $\beta^{opt}_3$   &Match Ratio  & $V_n(\vbetah_n)$\\ 
		\hline
		\multicolumn{7}{c}{Setting 1}\vspace{0.5em}\\   
		300&0.03 (0.33) &  0 (0) &  0.03 (0.30) &  0.03 (0.35) & 99.60\% &  0.00 (0.17) \\ 
		500&0.01 (0.21) &   0 (0) &  0.02 (0.20) &  0.02 (0.22) & 99.75\% &  0.00 (0.13)\\
		1000 &0.02 (0.13) &  0 (0) &  0.01 (0.13) &  0.01 (0.15) & 99.73\% & -0.01 (0.09) \\ 
		\hline 
		\multicolumn{7}{c}{Setting 2}\vspace{0.5em}	\\ 
		300&-0.04 (0.25) &  0 (0) &  0.02 (0.24) &  0.00 (0.17) & 99.40\% & -0.01 (0.15)\\ 
		500& -0.03 (0.20) & 0 (0) &  0.02 (0.19) &  0.00 (0.13) & 99.58\% & -0.01 (0.12)\\
		1000&0.00 (0.13) & 0 (0) &  0.01 (0.13) &  0.00 (0.09) & 99.90\% &  0.00 (0.08) \\
		\hline
	\end{tabular}  \vskip 18pt
\end{table} 

\begin{table}[!ht]   
	\def~{\hphantom{0}} 
	\caption{Empirical coverage probabilities and average interval lengths of the 95\% bootstrap confidence intervals for $\vbeta^{opt}$ with binary outcomes.}
	\centering
	\label{table: binary_inf} 
	\begin{tabular}{ccccccc}
		\hline
		$n$ & &$\beta^{opt}_0$&$\beta^{opt}_1$  &  $\beta^{opt}_2$  &  $\beta^{opt}_3$  & $V(\vbeta^{opt})$\\
		\hline
		\multicolumn{7}{c}{Setting 1} \vspace{0.5em}\\ 
		\multirow{2}{*}{300}  &Coverage Rate&91.1\% & 100\% & 91.2\% & 90.5\%& 94.5\% \\
		& Average Length&1.35 & 0& 1.24 & 1.38  & 0.69\\ 
		\multirow{2}{*}{500}  & Coverage Rate&94.6\% & 100\% & 93.1\% & 93.3\%   & 94.3\%\\ 
		&  Average Length&0.85 & 0 & 0.82 & 0.88& 0.54\\
		\multirow{2}{*}{1000}  & Coverage Rate&94.6\% & 100\% & 95.1\% & 93.8\% & 96.0\%\\ 
		& Average Length&0.56 & 0 & 0.55 & 0.58& 0.38\\ 
		\hline
		\multicolumn{7}{c}{Setting 2}\vspace{0.5em} \\ 
		\multirow{2}{*}{300}  &Coverage Rate &93.5\% & 100\%& 93.8\% & 97.6\%  & 94.4\%\\ 
		&  Average Length&1.11 & 0 & 1.03 & 0.72& 0.62\\ 
		\multirow{2}{*}{500}  & Coverage Rate&92.6\% & 100\% & 94.3\% & 96.0\% & 93.7\%\\ 
		& Average Length &0.78 & 0& 0.74 & 0.53  & 0.48\\ 
		\multirow{2}{*}{1000}  &Coverage Rate &94.4\% & 100\% & 94.6\% & 95.7\%   & 93.6\%\\ 
		& Average Length&0.52 & 0 & 0.50 & 0.36 &0.34 \\\hline
	\end{tabular}\vskip 18pt
\end{table}

\noindent {\bf Example 2} (different choices of kernel function). 
We consider the same data generative model as in Settings 1 \& 2 in the main paper, and evaluate two different choices of kernels $K(\cdot)$. 
The first choice uses $K_1(\cdot)=\Phi(\cdot)$, the cumulative distribution function of standard normal distribution.
The second choice is $K_2(v) =\big[0.5 + \frac{105}{64}\{\frac{v}{5}-\frac{5}{3}(\frac{v}{5})^3 +\frac{7}{5}(\frac{v}{5})^5 - \frac{3}{7}(\frac{v}{5})^7\}\big]I( -5\leq v \leq 5)+I(v>5)$. Its bandwidth is selected by  $h_n=0.9n^{-1/9} \min\{\mbox{ std} (\vx_i^T\vbeta),\mbox{ IQR}(\vx_i^T\vbeta)/1.34\}$.
Both choices satisfy the regularity conditions in the paper. The bandwidth was chosen the same way as described in 
Section 4 of the main paper. The simulation results are summarized in
Table~\ref{table: kernels}. We observe that the performance is not sensitive to different choices of kernel functions.

\begin{table}[!ht]
	\centering
	\def\~{\hphantom{0}}
	\caption{Monte Carlo estimates of 
		the bias and standard deviation of the estimate for the parameters indexing the optimal treatment regime, the match ratio (percentage of times the estimated optimal treatment regime matches the theoretically optimal treatment regime), and the bias and standard deviation of the estimated optimal value with with different choices of $K(\cdot)$.}
	\label{table: kernels}
	\begin{tabular} {cccccccc} 
		\hline 
		$n$ & Kernel&$\beta^{opt}_0$&$\beta^{opt}_1$  &  $\beta^{opt}_2$  &  $\beta^{opt}_3$   &Match Ratio  & $V_n(\vbetah_n)$\\ 
		\hline
		\multicolumn{8}{c}{Setting 1}\vspace{0.5em}\\   
		\multirow{2}{*}{300}
		&$K_1$ & -0.05 (0.30) &  0 (0) &  0.01 (0.27) &  0.04 (0.31) & 99.35\% & -0.02 (0.17)   \\ 
		&$K_2$ &-0.05 (0.27) &  0 (0) &  0.04 (0.26) &  0.05 (0.28) & 99.40\% & -0.01 (0.16)  \\ 
		\multirow{2}{*}{500} &$K_1$ &  -0.01 (0.19) &  0 (0)&  0.01 (0.20) &  0.02 (0.22) & 99.73\% &  0.00 (0.13) \\ 
		&$K_2$ &-0.03 (0.21) &  0 (0) &  0.02 (0.20) &  0.03 (0.22) & 99.60\% & -0.01 (0.13)\\  
		\multirow{2}{*}{1000} 
		&$K_1$  & -0.01 (0.14) & 0 (0) &  0.00 (0.13) &  0.01 (0.15) & 99.88\% & -0.01 (0.09)  \\ 
		&$K_2$ & -0.01 (0.14) &   0 (0) &  0.02 (0.14) &  0.02 (0.14) & 99.77\% &  0.00 (0.09)\\  
		\hline 
		\multicolumn{8}{c}{Setting 2}\vspace{0.5em}	\\ 
		\multirow{2}{*}{300} 
		&$K_1$ &0.04 (0.26) &  0 (0)  &  0.02 (0.24) &  0.02 (0.18) & 99.35\% & -0.01 (0.15) \\ 
		&$K_2$ &-0.03 (0.24) &   0 (0)   &  0.02 (0.24) &  0.00 (0.17) & 99.47\% & -0.01 (0.14)\\  
		\multirow{2}{*}{500} 
		& $K_1$& 0.02 (0.19) &  0 (0)&  0.02 (0.18) &  0.00 (0.13) & 99.65\% & -0.01 (0.11) \\ 
		&$K_2$ & -0.03 (0.20) &  0 (0)&  0.02 (0.19) &  0.01 (0.13) & 99.58\% & -0.01 (0.12)\\  
		\multirow{2}{*}{ 1000} 
		& $K_1$ &  -0.01 (0.14) & 0 (0) &  0.01 (0.13) &  0.00 (0.09) & 99.79\% & -0.01 (0.08)    \\
		& $K_2$ &  0.01 (0.13) & 0 (0) &  0.00 (0.13) &  0.00 (0.10) & 99.79\% & -0.01 (0.08)  \\ 
		\hline
	\end{tabular}  \vskip 18pt
\end{table}

\noindent {\bf Example 3} (observational data).  The data generative model is the same as that in in Settings 1 \& 2 in the main paper, except that 
$A$ is generated according to $P(A=1|\vx) = \{1+\exp(-\vx^T\veta)\}^{-1}$, where $\veta = (0.2,0.5,0.5,0.5)^T$.
Table~\ref{table: obs} summarizes the performance of the propensity score inverse weighted estimator given in
(\ref{obs}) of the main paper. We observed satisfactory performance.

\begin{table}[!ht]
	\centering
	\def\~{\hphantom{0}}
	\caption{Monte Carlo estimates of 
		the bias and standard deviation of the estimate for the parameters indexing the optimal treatment regime, the match ratio (percentage of times the estimated optimal treatment regime matches the theoretically optimal treatment regime), and the bias and standard deviation of the estimated optimal value in an observational study.}
	\label{table: obs}
	\begin{tabular} {ccccccc} 
		\hline 
		$n$ &$\beta^{opt}_0$&$\beta^{opt}_1$  &  $\beta^{opt}_2$  &  $\beta^{opt}_3$   &Match Ratio  & $V_n(\vbetah_n)$\\ 
		\hline
		\multicolumn{7}{c}{Setting 1}\vspace{0.5em}\\   
		300&-0.05 (0.28) & 0 (0) &  0.03 (0.32) &  0.04 (0.33) & 99.40\% & -0.01 (0.15) \\ 
		500&-0.03 (0.23) &  0 (0) &  0.03 (0.29) &  0.03 (0.27) & 99.63\% & -0.01 (0.12) \\
		1000 &-0.03 (0.15) &  0 (0) &  0.03 (0.18) &  0.03 (0.18) & 99.59\% & -0.01 (0.08) \\ 
		\hline 
		\multicolumn{7}{c}{Setting 2}\vspace{0.5em}	\\ 
		300&-0.06 (0.27) &  0 (0) &  0.03 (0.29) &  0.01 (0.19) & 99.01\% & -0.01 (0.15)\\ 
		500&-0.02 (0.20) & 0 (0)&  0.02 (0.23) &  0.01 (0.15) & 99.66\% & -0.01 (0.12)\\
		1000&-0.02 (0.14) & 0 (0)&  0.02 (0.15) &  0.01 (0.10) & 99.69\% & -0.01 (0.09) \\
		\hline
	\end{tabular}  \vskip 18pt
\end{table}

\noindent {\bf Example 4} (addition results fro real-data example).  Table~\ref{table: real} shows the smooth and nonsmooth estimators for the real example in Section~\ref{sec:realdata}, with a 5-fold cross-validation.
Specifically, we randomly divide the data into five folds and use four folds to estimate $\vbeta_0$ and $V(\beta_0)$ and evaluate
the matching ratio on the remaining fold (i.e., validation data). Each of the five folds is used as validation data in turn (refereed to as iterative 1, \ldots, 5 in te table), the final results
are summarized in 	Table~\ref{table: real}.
We observe that the smooth estimators always lie in the element-wise confidence intervals we calculated in Section~\ref{sec:realdata}. 
However, the nonsmooth estimators are rather nonstable and can even change signs across iteratives.

\begin{table}[!ht]
	\centering
	\def\~{\hphantom{0}}
	\caption{Real data example: comparison of smooth and nonsmooth estimators based on 5-fold cross-validation}
	\label{table: real}
	\begin{tabular} {cccccc} 
		\hline 
		Iterative & Method&$\beta^{opt}_0$&$\beta^{opt}_1$  &  $\beta^{opt}_2$  &Match Ratio  \\ 
		\hline 
		\multirow{2}{*}{1}
		& Smooth&0.65 & 1 &  0.46 & \multirow{2}{*}{87.93\%} \\ 
		& Nonsmooth&-0.40 &1&  0.36 &   \\ 
		\multirow{2}{*}{2}
		& Smooth&0.45 & 1 &  0.35 & \multirow{2}{*}{86.21\%} \\ 
		& Nonsmooth&-0.39 &1&  -0.35 &   \\ 
		\multirow{2}{*}{3}
		& Smooth&0.64 & 1 &  0.42 & \multirow{2}{*}{89.47\%} \\ 
		& Nonsmooth&-0.42 &1&  -0.16 &   \\ 
		\multirow{2}{*}{4}
		& Smooth&0.46 & 1 &  0.27 & \multirow{2}{*}{85.96\%} \\ 
		& Nonsmooth&0.14 &1&  -0.73 &   \\ 
		\multirow{2}{*}{5}
		& Smooth&0.64 & 1 &  0.46 & \multirow{2}{*}{87.72\%}\\ 
		& Nonsmooth&-0.40 &1&  13.38 &  \\ 
		\hline
	\end{tabular}  \vskip 18pt
\end{table}

\end{document}